\documentclass[useAMS,usenatbib]{mn2e}
\usepackage{graphicx}
\usepackage{natbib}

\newcommand{\ltsima} {$\; \buildrel < \over \sim \;$}
\newcommand{\gtsima} {$\; \buildrel > \over \sim \;$}
\newcommand{\lta} {\lower.5ex\hbox{\ltsima}}
\newcommand{\gta} {\lower.5ex\hbox{\gtsima}}

\title{WMAP 5-year constraints on $f_{nl}$ with wavelets}

\author[A. Curto et al.]{A. Curto,$^1$ $^2$\thanks{e-mail:
curto@ifca.unican.es} E. Mart\'{\i}nez-Gonz\'alez,$^1$ P. Mukherjee,$^3$ R. B. Barreiro,$^1$  
\newauthor F. K. Hansen,$^4$  M. Liguori,$^5$  S. Matarrese$^6$ \\ 
$^1$     Instituto de F\'isica de Cantabria, CSIC-Universidad de Cantabria, Avda. de los Castros s/n, 39005 Santander, Spain.\\
$^2$     Dpto. de F\'isica Moderna, Universidad de Cantabria, Avda. los Castros s/n, 39005 Santander, Spain. \\
$^3$     Astronomy Centre, University of Sussex, Brighton BN1 9QH, United Kingdom. \\
$^4$     Institute of Theoretical Astrophysics, University of Oslo, P.O. Box 1029 Blindern, 0315 Oslo, Norway. \\
$^5$     Department of Applied Mathematics and Theoretical Physics, Centre for Mathematical Sciences, University of Cambridge, \\
~~Wilberfoce Road, Cambridge, CB3 0WA, United Kingdom. \\
$^6$  Dipartimento di Fisica “G. Galilei”, Universit\`a di Padova and INFN, Sezione di Padova, via Marzolo 8, I-35131, Padova, Italy}

\date{Accepted  Received ; in original form }

\begin{document}

\maketitle

\begin{abstract}
We present a Gaussianity analysis of the WMAP 5-year Cosmic Microwave
Background (CMB) temperature anisotropy data maps. We use several
third order estimators based on the spherical Mexican hat wavelet.  We
impose constraints on the local non-linear coupling parameter $f_{nl}$
using well motivated non-Gaussian simulations. We analyse the WMAP
maps at resolution of 6.9 arcmin for the Q, V, and W frequency
bands. We use the $KQ$75 mask recommended by the WMAP team which masks
out 28\% of the sky. The wavelet coefficients are evaluated at 10
different scales from 6.9 to 150 arcmin. With these coefficients we
compute the third order estimators which are used to perform a
$\chi^2$ analysis. The $\chi^2$ statistic is used to test the
Gaussianity of the WMAP data as well as to constrain the $f_{nl}$
parameter. Our results indicate that the WMAP data are compatible with
the Gaussian simulations, and the $f_{nl}$ parameter is constrained to
$-8 < f_{nl} < +111$ at 95\% CL for the combined V+W map. This value
has been corrected for the presence of undetected point sources, which
add a positive contribution of $\Delta f_{nl}=3 \pm 5$ in the V+W
map. Our results are very similar to those obtained by
\citet{komatsu2008} using the bispectrum.
\end{abstract}
\section{Introduction}
A valuable source of information of the early universe is the cosmic
microwave background (CMB) radiation. The temperature fluctuations of
this radiation can be used to put tight constraints on the
cosmological parameters. The Big Bang theory and the inflationary
models are our best theories to describe the universe. In particular,
the standard, single field, slow roll inflation
\citep{guth,albrecht,linde1982,linde1983} is one of the most accepted
models because of the accuracy of its predictions. In particular, it
forecasts that the primordial density fluctuations are compatible with
a nearly Gaussian random field. However, different primordial
processes (as for example topological defects, etc.) can introduce
non-Gaussian features at certain levels that may be detected
\citep{cruzc}. Also, several non-standard models of inflation predict
a detectable level of non-Gaussianity in the primordial gravitational
potential \citep[see, e.g.][and refs. therein]{bartolo}. Other
non-Gaussian deviations can be explained by the presence of foreground
contamination or systematic errors. In any case, the search of
non-Gaussian deviations in the CMB has become a question of
considerable interest, as it can be used to discriminate different
possible scenarios of the early universe and also to study the
secondary sources of non-Gaussianity.

Many tests of Gaussianity have been performed on different CMB data
sets. For a recent review of this topic see for example
\citet{martinez2008}. Some of the tests have found that the data are
compatible with Gaussianity but there are some important detections of
non-Gaussian deviations \citep[see for
  example][]{eriksen2004,eriksen2005,copi2004,copi2006,vielva2004,cruza,wiaux2006,vielva2007,monteserin2008},
all of them using the {\it Wilkinson Microwave Anisotropy
  Probe}\footnote{http://map.gsfc.nasa.gov/} (WMAP) data.

High precision experiments as the WMAP are sensitive to deviations due
to second-order effects in perturbation theory, usually parametrised
through the local non-linear coupling parameter $f_{nl}$. Several
methods have been used to constrain this parameter using the data of
different experiments. We can mention the angular bispectrum on WMAP
data \citep{komatsu2003,creminelli2006,spergel,komatsu2008} and COBE
\citep{komatsu2002}; the Minkowski functionals on WMAP
\citep{komatsu2003,spergel,gott2007,hikage2008,komatsu2008}, BOOMERanG
\citep{troia} and Archeops \citep{curto2007,curto2008}; different kind
of wavelets on WMAP \citep{mukherjee2004,cabella2005} and COBE
\citep{cayon2003} among others. The previous works suggest that
$f_{nl}$ is compatible with zero at least at 95\% confidence
level. However other recent works \citep{smoot2007,wandelt2008}
suggest a positive detection of $f_{nl}$ at a confidence level greater
than 95\% using WMAP data.

In this paper we perform a wavelet-based analysis of the 5-year
WMAP data in order to constrain the $f_{nl}$ parameter.
We use the high resolution WMAP data maps (6.9 arcmin) and realistic
non-Gaussian simulations performed following the algorithms developed
by \citet{liguori2003,liguori2007}.

Our article is organised as follows. Section \ref{methodology}
presents the statistical method, the estimators that we use to test
Gaussianity and constrain $f_{nl}$, as well as the data maps and
simulations to be analysed. In Section \ref{results} we summarise the
results of this work. The conclusions are drawn in Section
\ref{conclusions}.
\section{Methodology}
\label{methodology}
\subsection{The SMHW}
For this analysis we use a wavelet-based technique. The considered
wavelet is the Spherical Mexican Hat Wavelet (SMHW) as defined in
\citet{smhw}. The spherical wavelets have been used in some analyses
to test the Gaussianity of different data sets. We can mention the
analysis of the COBE-DMR data \citep{barreiro2000,cayon2001,cayon2003}
and WMAP data
\citep{vielva2004,vielva2007,mukherjee2004,cabella2005,cayon2005,mcewen2005,cruza,cruzb,wiaux2008}
among others. The SMHW can be obtained from the Mexican hat wavelet in
the plane $R^2$ through a stereographic projection
\citep{antoine1998}. Given a function $f({\bf n})$ evaluated on the
sphere at a direction ${\bf n}$ and a continuous wavelet family on
that space $\Psi({\bf n}; {\bf b}, R)$, we define the continuous
wavelet transform as
\begin{equation}
w({\bf b}; R) = \int d{\bf n}f({\bf n})\Psi({\bf n}; {\bf b}, R)
\label{wavmap}
\end{equation}
where ${\bf b}$ is the position on the sky at which the wavelet
coefficient is evaluated and $R$ is the scale of the wavelet.  In the
case of the SMHW we have that the wavelet only depends on the polar
angle $\theta$ and the scale R \citep{smhw}. In particular we have
\begin{equation}
\Psi_S(\theta; R) = \frac{1}{\sqrt{2\pi}N(R)}\left [ 1+ \left( \frac{y}{2}\right)^2\right ]^2\left [ 2 - \left( \frac{y}{R}\right)^2\right ]e^{-y^2/2R^2}
\end{equation}
where 
\begin{equation}
N(R)=R\left(1+\frac{R^2}{2}+\frac{R^4}{4}\right)^{1/2}
\end{equation}
and
\begin{equation}
y = 2\tan\left( \frac{\theta}{2}\right).
\end{equation}
We can compute the wavelet coefficient map given by Eq. \ref{wavmap}
at several different scales $R_j$ in order to enhance the non-Gaussian
features dominant at a given scale. In particular, we will consider 9
scales: $R_1=6.9'$, $R_2=10.3'$, $R_3=13.7'$, $R_4=25'$, $R_5=32'$,
$R_6=50'$, $R_7=75'$, $R_8=100'$, and $R_9=150'$. In addition we will
consider the unconvolved map, which will be represented hereafter as
scale $R_0$.  Larger scales are less sensitive to the local $f_{nl}$
model.
\label{scales}
\subsection{The estimators}
\label{theestimators}
In previous works, constraints on $f_{nl}$ with wavelets have been
obtained using only the skewness of the wavelet coefficients at
different scales \citep[e.g.][]{mukherjee2004,cabella2005}. In the
present work, other third order moments involving different scales are
also considered\footnote{Notice that inter-scale wavelet
    estimators have been previously used in the context of blind
    Gaussianity analyses, see e.g. \citet{pando1998,mukherjee2000}
    (for analyses involving two scales) and \citet{cayon2001}
    (involving three scales).}. As it will be shown in the results
section, the combination of these estimators is as efficient as the
bispectrum.  The estimators that we use in this analysis are based on
third order combinations of the wavelet coefficient maps $w_i(R_j)$
evaluated in sets of three contiguous scales. For each scale $R_j$ and
the next two scales $R_{j+1}$ and $R_{j+2}$ we can define
\begin{eqnarray}
\nonumber
q_1(R_j) &=& \frac{1}{N_j}\sum_{i=0}^{N_{pix}-1}\frac{w_{i,j}^3}{\sigma_j^3} \\
\label{wi3}
\nonumber
q_2(R_j,R_{j+1}) &=& \frac{1}{N_{j,j+1}}\sum_{i=0}^{N_{pix}-1}\frac{w_{i,j}^2w_{i,j+1}}{\sigma_j^2\sigma_{j+1}} \\
\label{wi2towip1}
\nonumber
q_3(R_j,R_{j+1}) &=& \frac{1}{N_{j,j+1}}\sum_{i=0}^{N_{pix}-1}\frac{w_{i,j}w_{i,j+1}^2}{\sigma_{j}\sigma_{j+1}^2} \\
\label{witowip12}
\nonumber
q_4(R_j,R_{j+1},R_{j+2}) &=& \frac{1}{N_{j,j+1,j+2}}\sum_{i=0}^{N_{pix}-1}\frac{w_{i,j}w_{i,j+1}w_{i,j+2}}{\sigma_{j}\sigma_{j+1}\sigma_{j+2}} \\
\label{witowip1towip2}
\nonumber
q_5(R_j,R_{j+2}) &=& \frac{1}{N_{j,j+2}}\sum_{i=0}^{N_{pix}-1}\frac{w_{i,j}^2w_{i,j+2}}{\sigma_{j}^2\sigma_{j+2}} \\
\label{wi2towip2}
\nonumber
q_6(R_j,R_{j+2}) &=& \frac{1}{N_{j,j+2}}\sum_{i=0}^{N_{pix}-1}\frac{w_{i,j}w_{i,j+2}^2}{\sigma_{j}\sigma_{j+2}^2} \\
\label{witowip22}
\end{eqnarray}
where $N_{j_1,j_2,j_3}$ is the number of available pixels after
combining the scales $R_{j_1}$, $R_{j_2}$, and $R_{j_3}$, $N_{pix}$ is
the total number of pixels, $w_{i,j}= w_i(R_j)$, and $\sigma_{j}$ is
the dispersion of $w_{i,j}$.  Each map $w_{i,j}$ is masked out with an
appropriate mask at the scale $R_j$ and its mean value outside the
mask is removed. 
These estimators have a Gaussian-like distribution when are computed
for a set of Gaussian simulations. Thus, we can use effectively a
$\chi^2$ statistics to test Gaussianity and to constrain
$f_{nl}$. Considering all the estimators evaluated in all the scales
we can construct a vector
\begin{equation}
{\bf v} = (q_1(R_0),q_2(R_0,R_1),q_3(R_0,R_1),,q_4(R_0,R_1,R_2),...)
\label{myvector}
\end{equation}
with a dimension of $n_v = n_{sc}+2(n_{sc}-1)+3(n_{sc}-2)$ for $n_{sc}
\ge 2$, where $n_{sc}$ is the number of considered scales. This vector
is used to compute the $\chi^2$ estimator
\begin{equation}
\chi^2=\sum_{k,l=0}^{n_v-1}(v_k - \langle v_k \rangle)C_{kl}^{-1}(v_l - \langle v_l \rangle)
\label{chigauss}
\end{equation}
where $\langle \rangle$ is the expected value for the Gaussian case
and $C_{kl}$ is the covariance matrix $C_{kl}=\langle v_kv_l \rangle -
\langle v_k \rangle \langle v_l \rangle$. The Gaussianity analysis
consists on computing the $\chi^2$ statistic for the data and compare
it to the distribution of this quantity obtained from Gaussian simulations.

The second part of the analysis consists on setting constraints on the
$f_{nl}$ through a $\chi^2$ test. In this case
\begin{equation}
\chi^2(f_{nl})=\sum_{k,l=0}^{n_v-1}(v_k - \langle v_k \rangle_{f_{nl}})C_{kl}^{-1}(f_{nl})(v_l - \langle v_l \rangle_{f_{nl}})
\label{chifnl}
\end{equation}
where $\langle \rangle_{f_{nl}}$ is the expected value for a model
with $f_{nl}$ and $C_{kl}=\langle v_kv_l \rangle_{f_{nl}} - \langle
v_k \rangle_{f_{nl}} \langle v_l \rangle_{f_{nl}}$. For low values of
$f_{nl}$ ($f_{nl} \leq 1500$) we can use the following approximation
$C_{kl}(f_{nl}) \simeq C_{kl}(f_{nl}=0) = C_{kl}$.  The best-fit
$f_{nl}$ for the data is obtained by minimization of
$\chi^2(f_{nl})$. Error bars for this parameter at different
confidence levels are computed using Gaussian simulations.
\subsection{Data and simulations}
\label{datasimu}
\begin{figure*}
\center
\includegraphics[width=3.7cm,height=6.4cm, angle = 90] {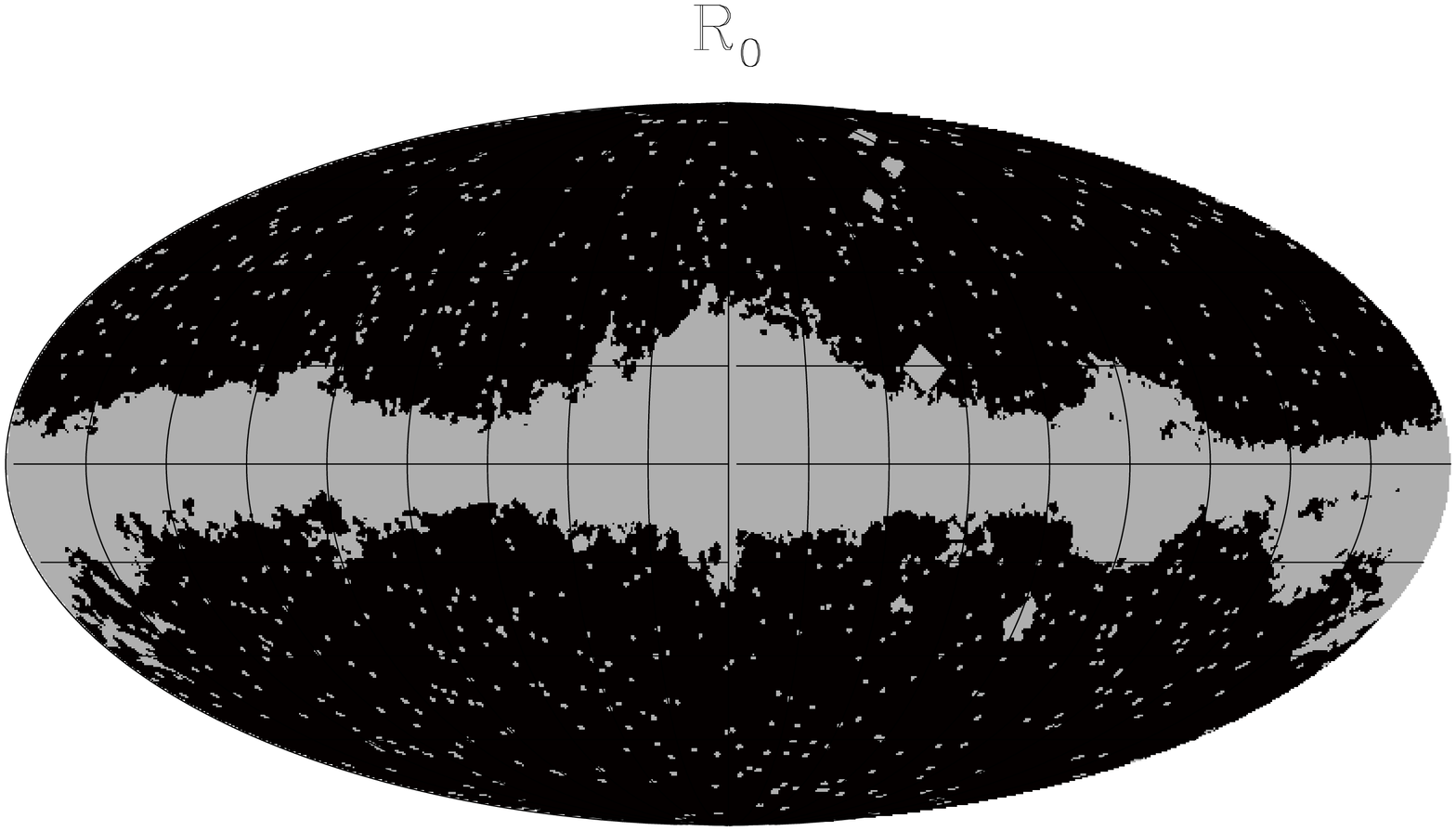}
%~~
\includegraphics[width=3.7cm,height=6.4cm, angle = 90] {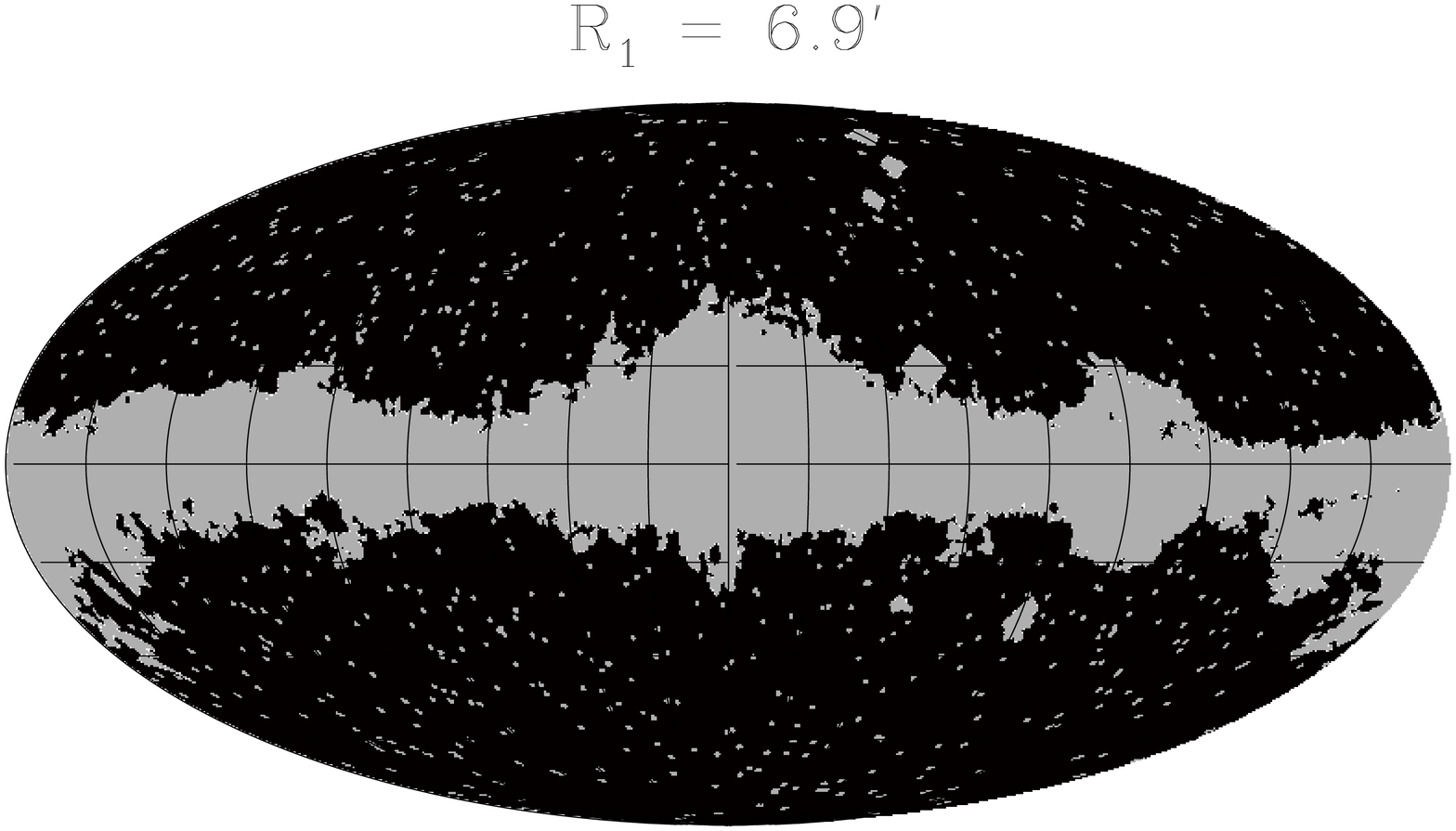}
%~~
\includegraphics[width=3.7cm,height=6.4cm, angle = 90] {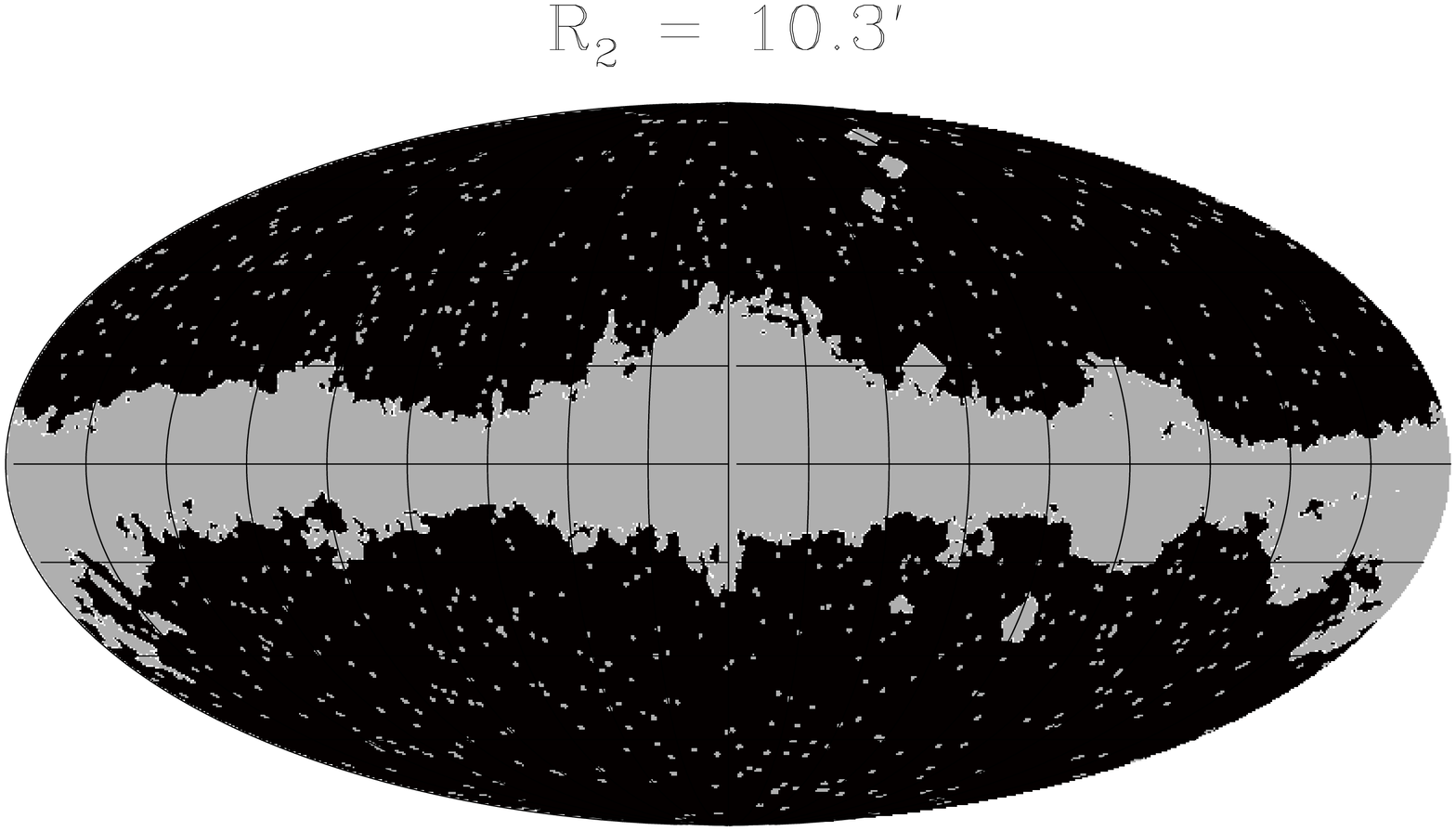}
%~~
\includegraphics[width=3.7cm,height=6.4cm, angle = 90] {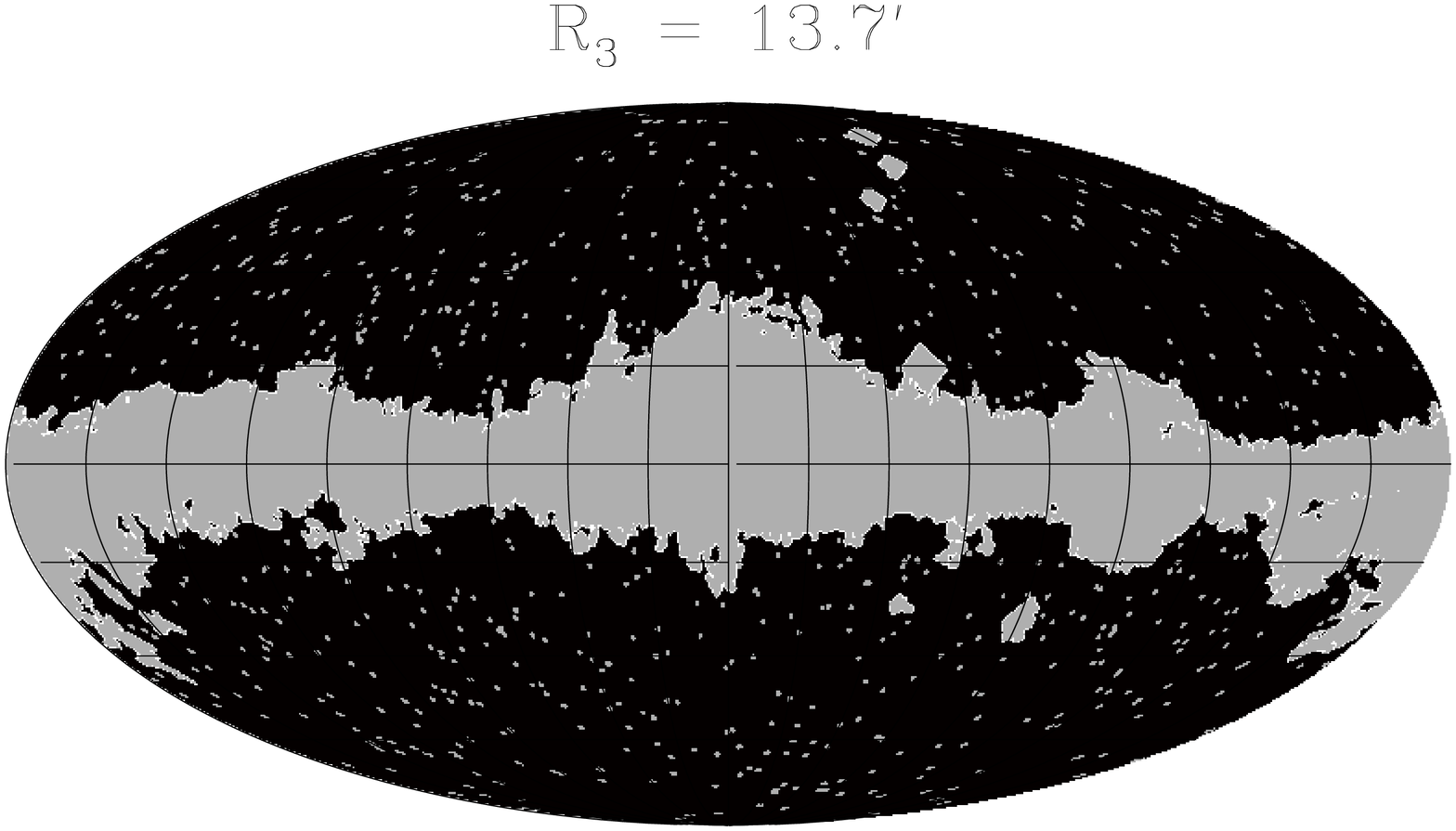}
%~~
\includegraphics[width=3.7cm,height=6.4cm, angle = 90] {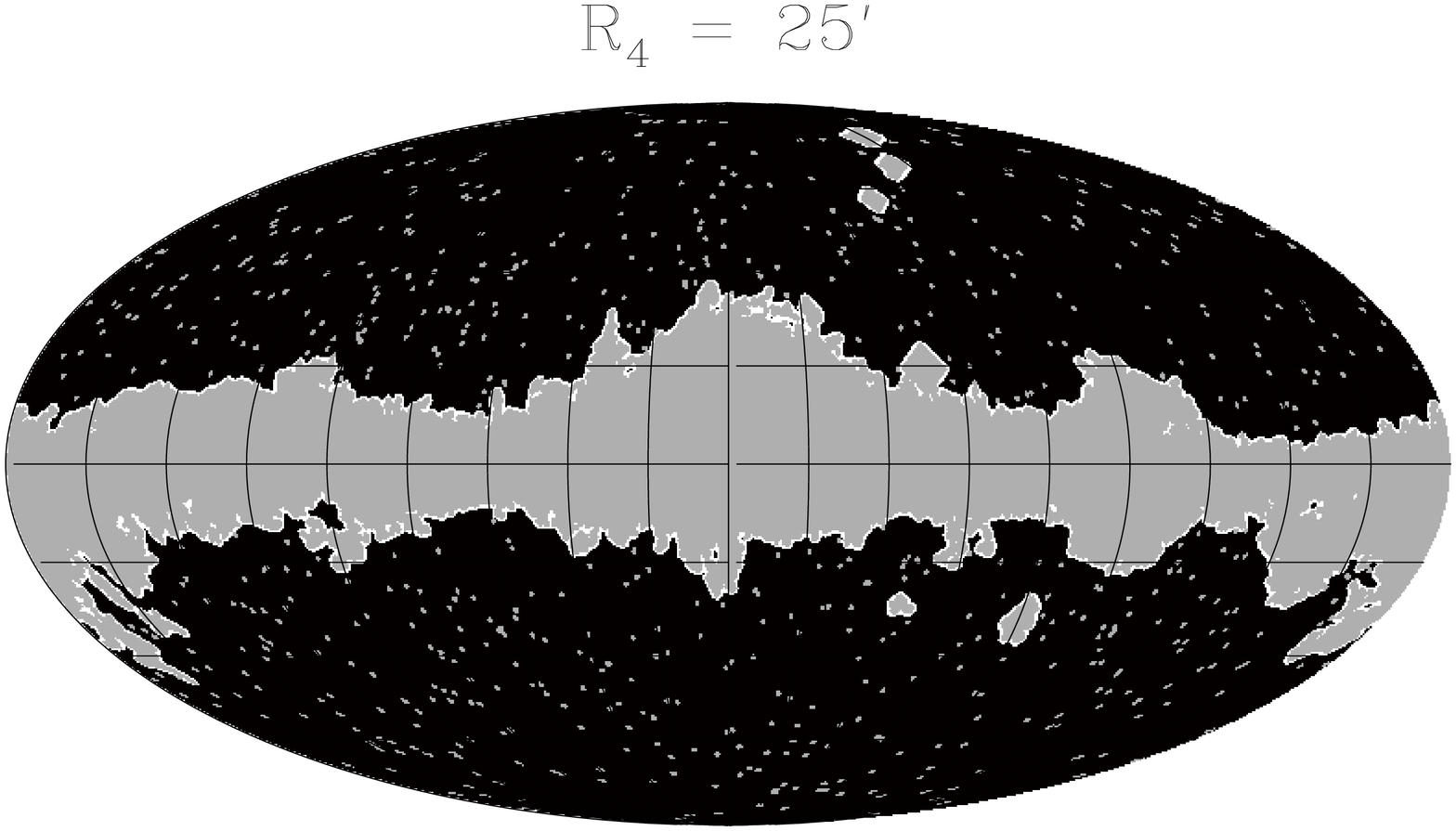}
%~~
\includegraphics[width=3.7cm,height=6.4cm, angle = 90] {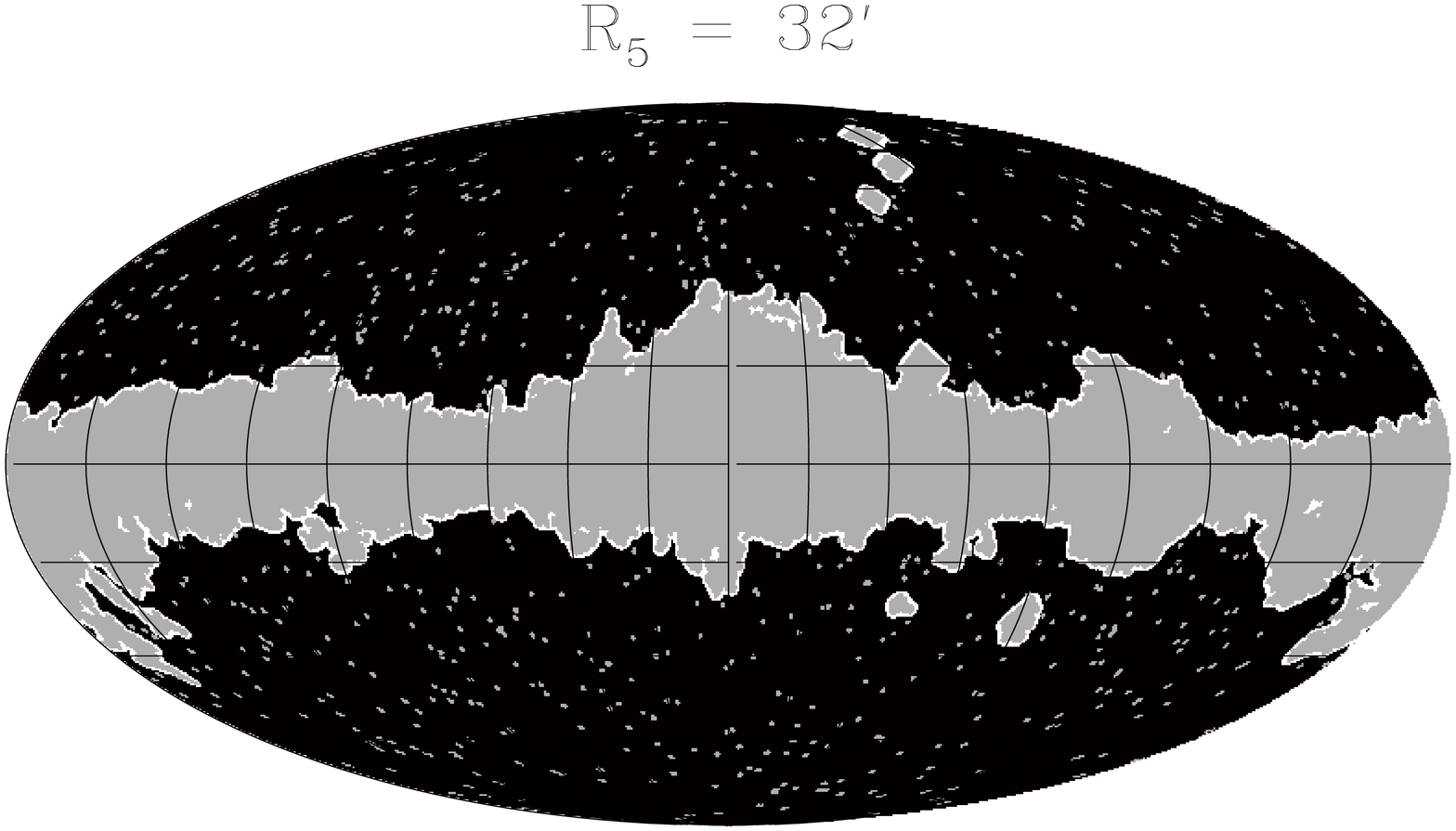}
%~~
\includegraphics[width=3.7cm,height=6.4cm, angle = 90] {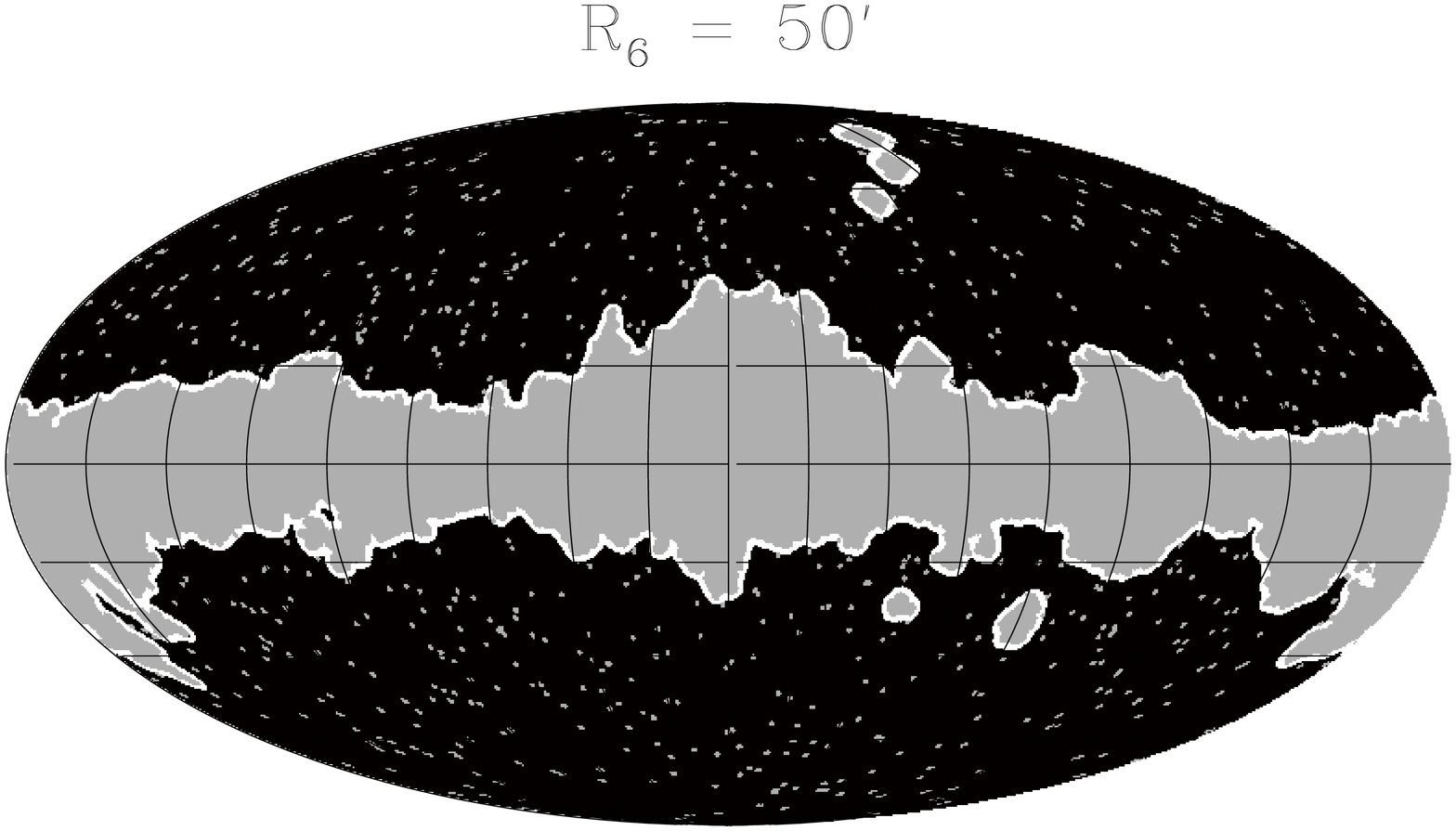}
%~~
\includegraphics[width=3.7cm,height=6.4cm, angle = 90] {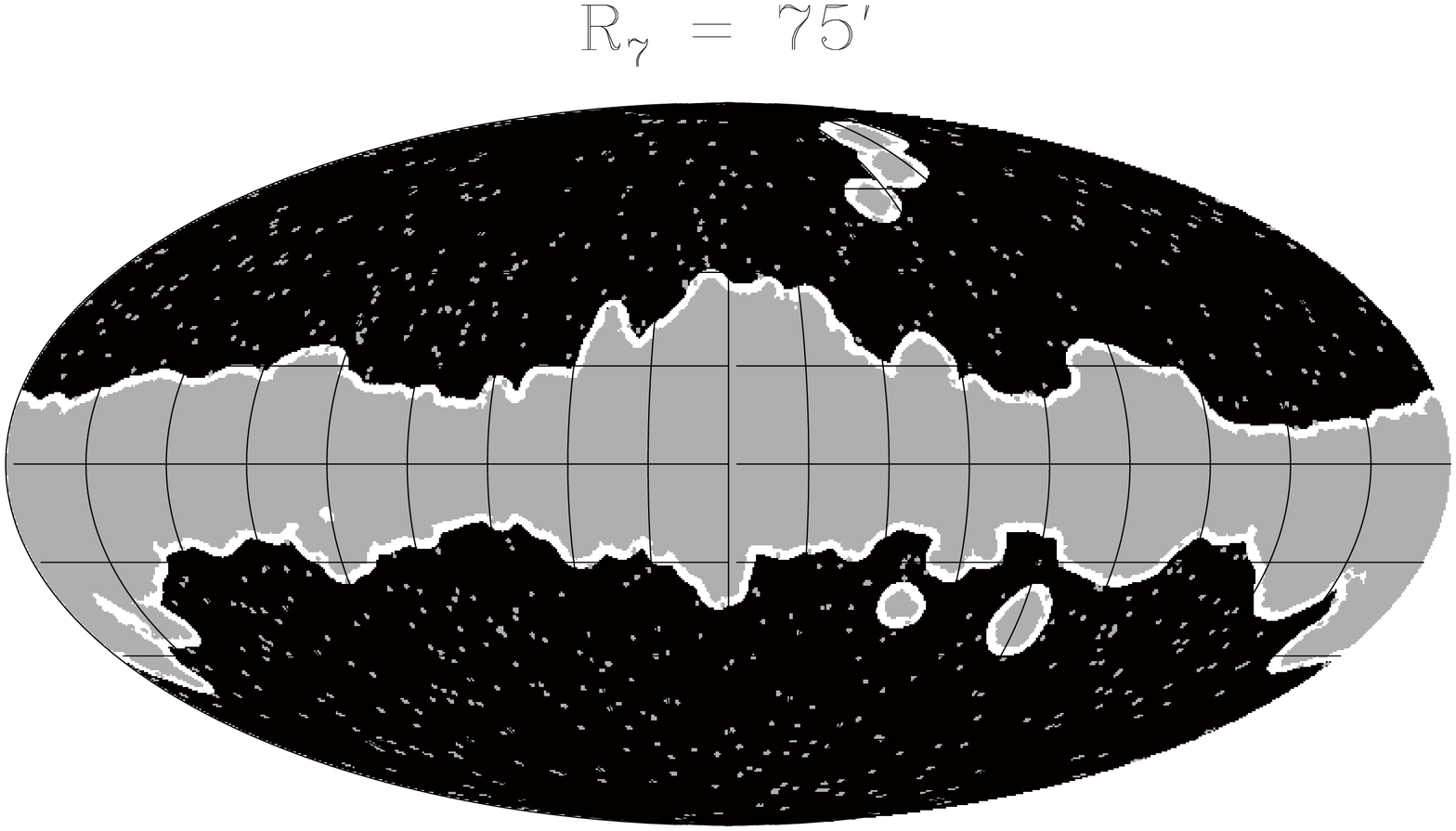}
%~~
\includegraphics[width=3.7cm,height=6.4cm, angle = 90] {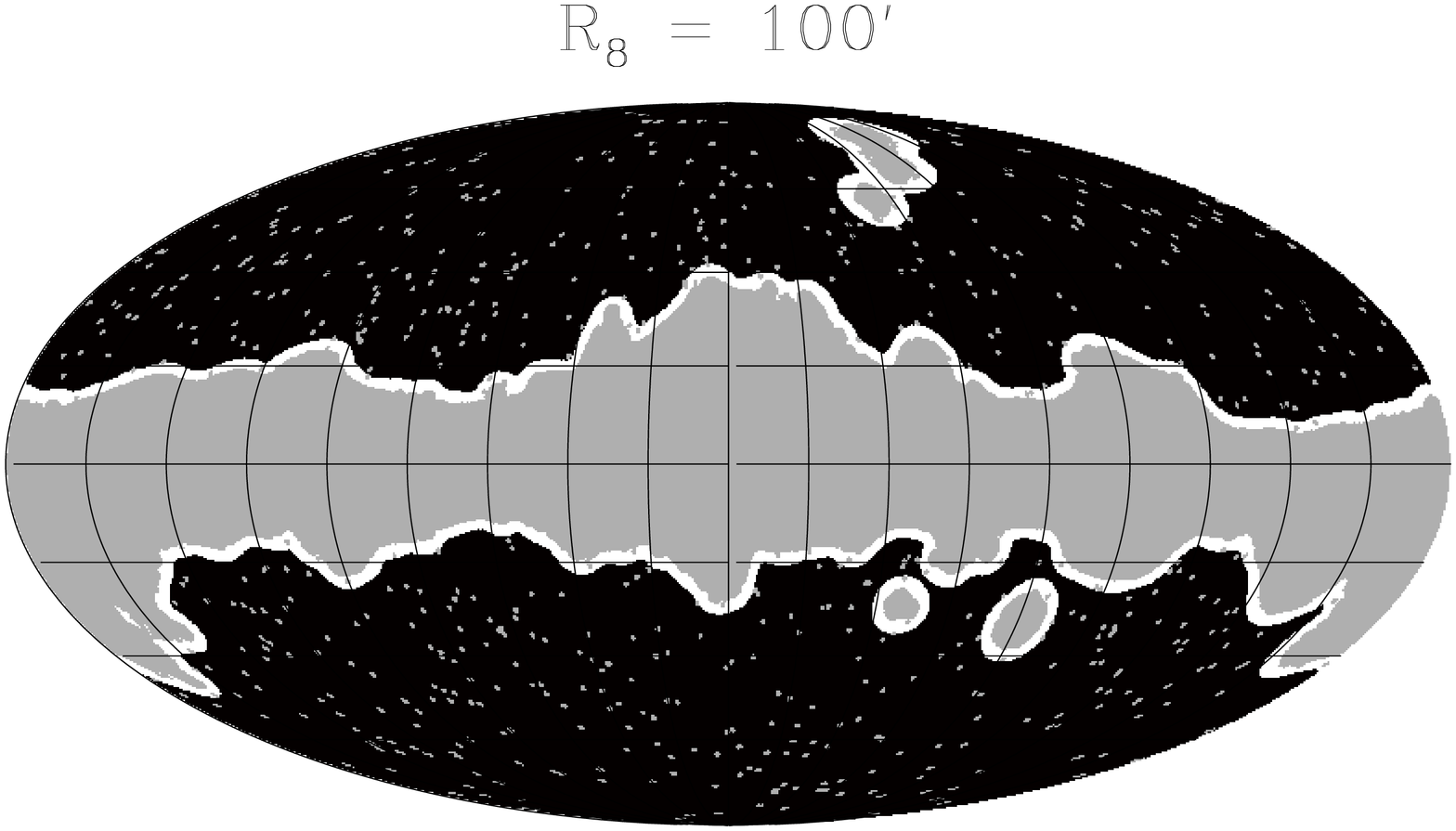}
%~~
\includegraphics[width=3.7cm,height=6.4cm, angle = 90] {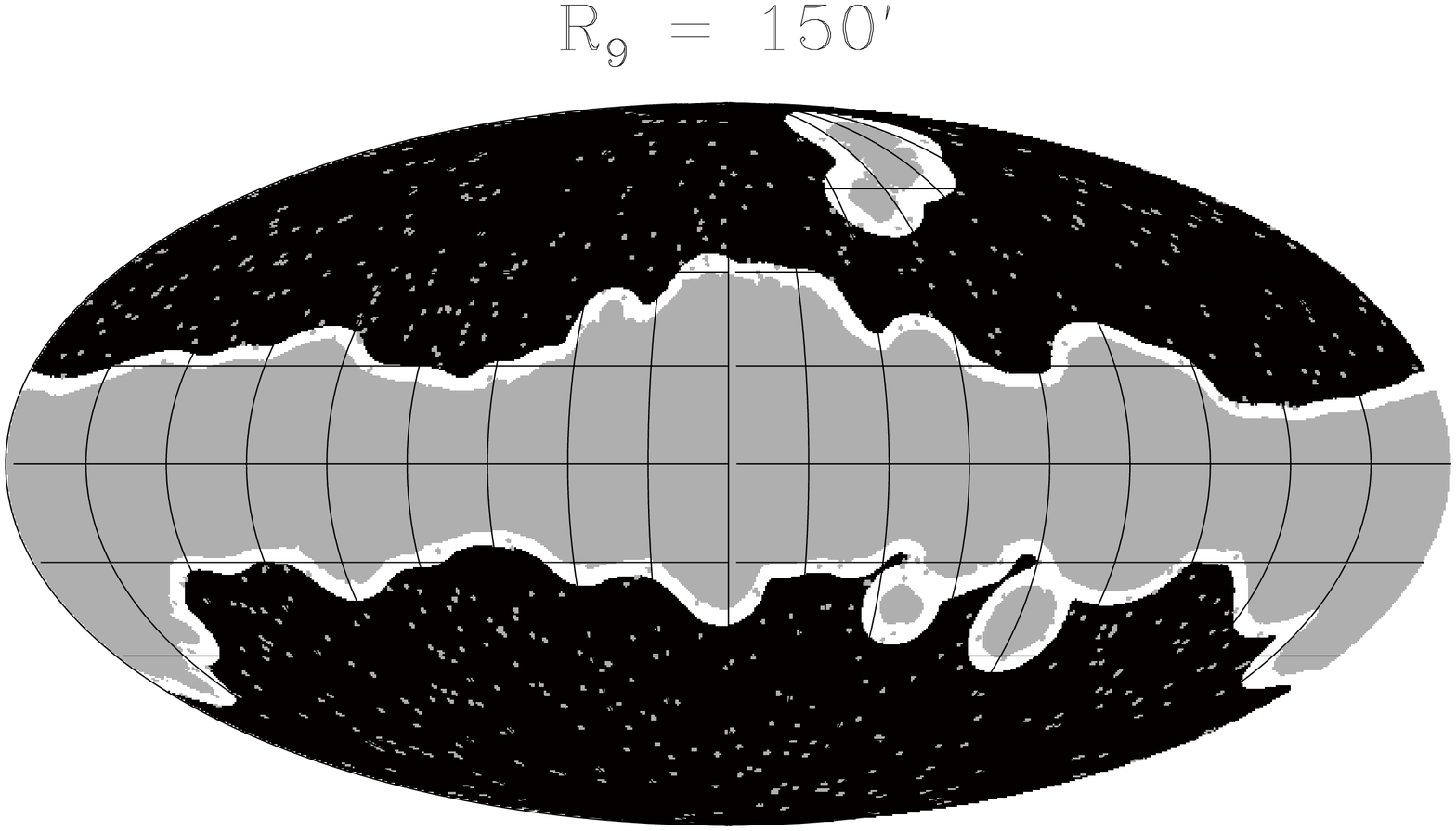}
%~~
\caption{Masks considered for the analysis at each wavelet scale. The
  case $R_0$ is the $KQ$75 mask. The others correspond to the
  restrictive extended masks (black) and the less restrictive extended
  masks (white+black). The grid corresponds to a size of 20 degrees.}
\label{masks}
\end{figure*}
For our analysis, we use the 5-year WMAP foreground reduced data,
which are available in the LAMBDA web
site\footnote{http://lambda.gsfc.nasa.gov}. We combine the maps of
different radiometers, using the inverse of the noise variance as an
optimal weight, as described in \citet{bennett2003}. In particular we
use five combined maps: Q+V+W, V+W, Q, V, and W. The Q+V+W map has the
eight maps corresponding to the two radiometers of the Q band (41 GHz),
two radiometers of the V band (61 GHz) and four radiometers of the W
band (94 GHz). Similar combinations are constructed for the V+W, Q, V
and W maps. The pixel resolution of these maps is 6.9 arcmin,
corresponding to a HEALPix \citep{healpix} $N_{side}$ parameter of
512. The mask that we use is the $KQ$75 which discards 28\% of the
sky.

We should take into account the effect of the mask in the wavelet
coefficient maps. Pixels near the border of the mask (the Galactic cut
and other features) are affected by the zero value of the mask
\citep{vielva2004}. Therefore we need one extended mask for each scale
that removes these affected pixels from the analysis. We use the
method described in \citet{mcewen2005} to construct our extended
masks. We compute the wavelet coefficients at each scale of the $KQ$75
mask without the holes corresponding to the point sources, and
consider only the pixels with low values (that is, the less affected
pixels). Then we multiply by $KQ$75 to mask out the point sources. In
particular, our threshold is 0.001, i.e., all the pixels which have a
wavelet coefficient for the $KQ$75 mask larger than 0.001 in absolute
value are masked out. In addition we also test the effect of the
  extension of the mask by applying a less restrictive threshold of
  0.01. In Fig. \ref{masks} we present the masks that we use for the
10 considered scales.

We also need Gaussian and non-Gaussian simulations for this
analysis. The Gaussian simulations are performed as follows. For a
given power spectrum $C_{\ell}$ (we use the best fit power spectrum
for WMAP provided by LAMBDA), we generate a set of Gaussian $a_{\ell
  m}$. From these multipoles we produce a map for each different
radiometer by convolving with the corresponding beam transfer
function. We also include the pixel properties by convolving with the
pixel transfer function. We add a Gaussian noise realisation to each
radiometer simulation and then we combine them in the same way as the
data maps. Following the analysis of the WMAP team, we assume
  that the instrumental noise is well approximated by Gaussian white
  noise at each pixel. This noise is characterised by a dispersion
  that depends on the pixel position and the corresponding
  radiometer. Although the data also contain small residuals of $1/f$
  noise \citep{jarosik2003,jarosik2007} their contribution here is
  expected to be negligible.

The non-Gaussian simulations are produced following a model which
introduces a quadratic term in the primordial gravitational potential
\citep{salopek,gangui,verde,komatsu2001}:
\begin{equation}
\Phi({\bf x}) = \Phi_L({\bf x}) + f_{nl}\{\Phi_L^2({\bf x})-\langle \Phi_L^2({\bf x}) \rangle \} \ \ 
\end{equation}
where $\Phi_L({\bf x})$ is a linear random field which is Gaussian
distributed and has zero mean. This kind of non-Gaussianity is
generated in various non-standard inflationary scenarios \citep[see,
  e.g.][]{bartolo}. The simulations with $f_{nl}$ are generated
following the algorithms described in
\citet{liguori2003,liguori2007}. In particular we have a set of 300
Gaussian simulations in the $a_{\ell m}$ space, $a_{\ell m}^{(G)}$ and
their corresponding non-Gaussian part $a_{\ell m}^{(NG)}$. A
simulation with a given value of $f_{nl}$ is constructed as
\begin{equation}
a_{\ell m} = a_{\ell m}^{(G)}+f_{nl}a_{\ell m}^{(NG)}.
\label{almfnl}
\end{equation}
To produce each non-Gaussian WMAP simulation with $f_{nl}$ we
transform the multipoles defined in Eq. \ref{almfnl} into each
differencing assembly map using the corresponding beam and pixel
transfer functions and we add a Gaussian noise realisation. Then, we
combine these maps to form the Q+V+W, V+W, Q, V and W combined maps.
\section{Results}
\label{results}
In this section we present the Gaussianity analysis of the WMAP data
using the Q+V+W, V+W, Q, V, and W combined data maps. First we
consider the estimators defined in Eqs. \ref{witowip22} and the
$\chi^2$ statistic given in Eq. \ref{chigauss} and compare the values
obtained for the WMAP data with the distribution constructed with
Gaussian simulations. Second we constrain the local non-linear
coupling parameter $f_{nl}$ using realistic non-Gaussian
simulations. Finally we estimate the contribution of the undetected
point sources to the best fit value of $f_{nl}$.
\subsection{Analysis of WMAP data}
\label{resultsfnl}
\begin{figure*}
\center
\includegraphics[width=7.0cm,height=4.4cm] {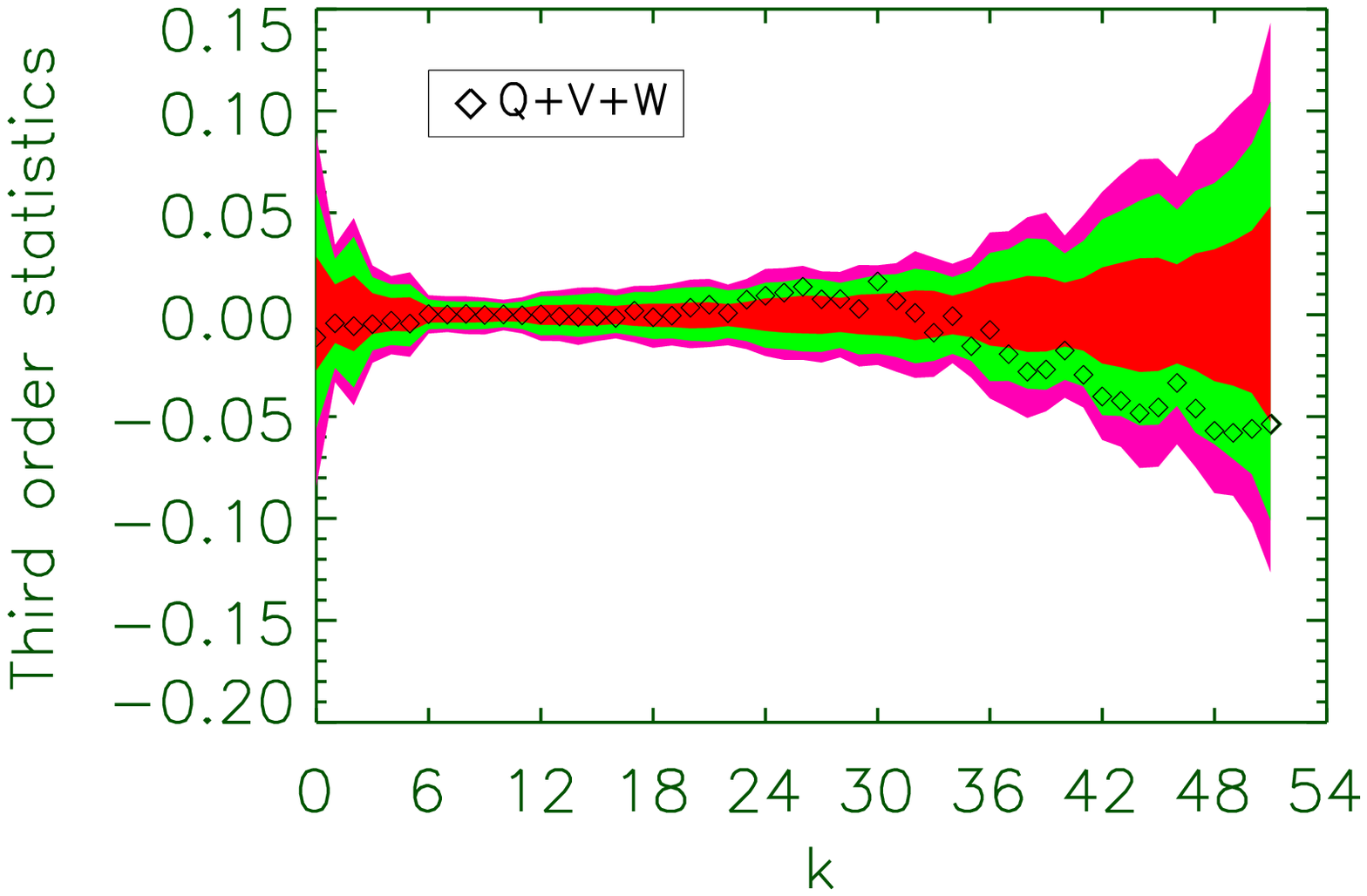}
%~~
\includegraphics[width=7.0cm,height=4.4cm] {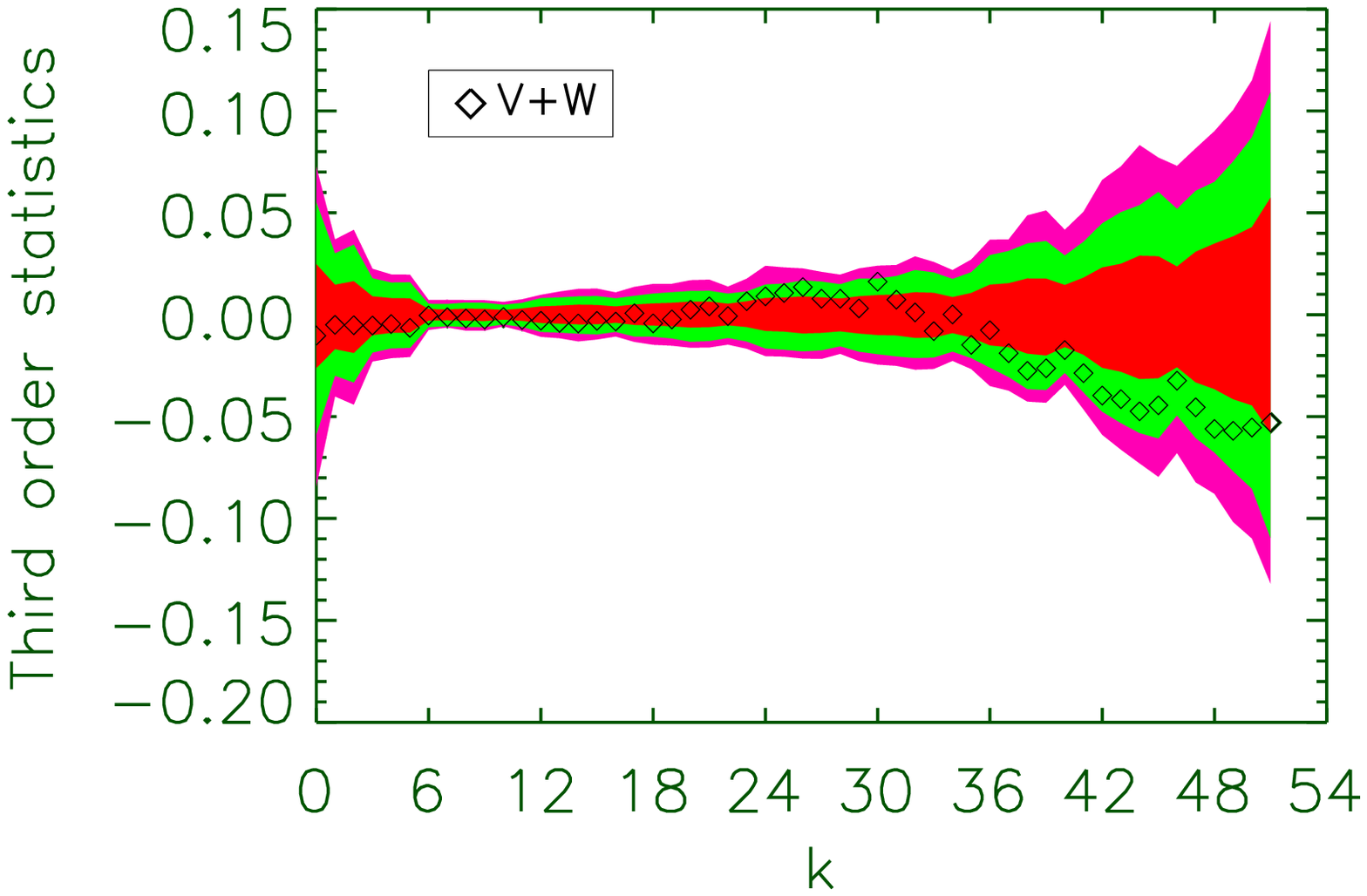}
%~~
\includegraphics[width=7.0cm,height=4.4cm] {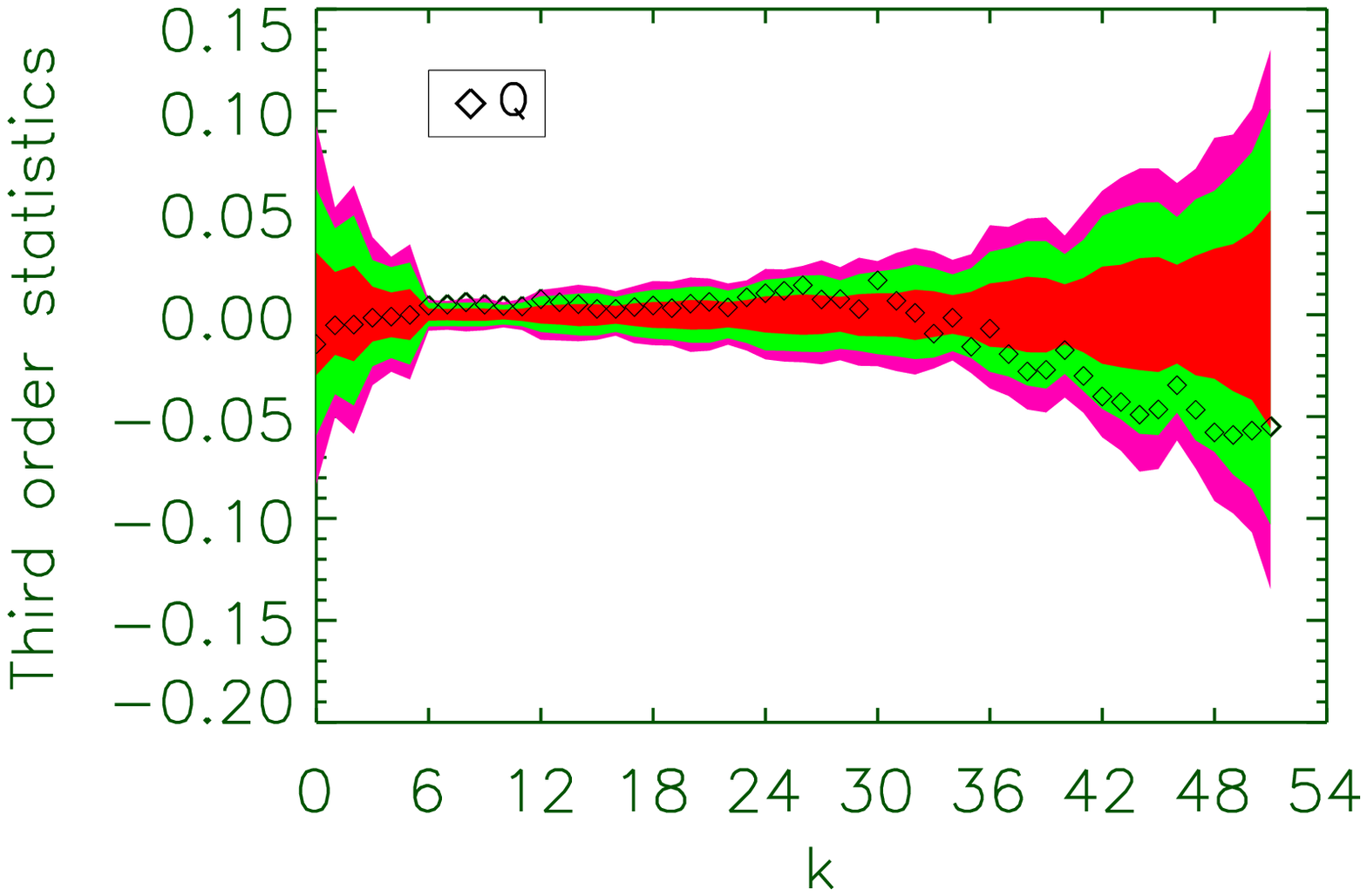}
%~~
\includegraphics[width=7.0cm,height=4.4cm] {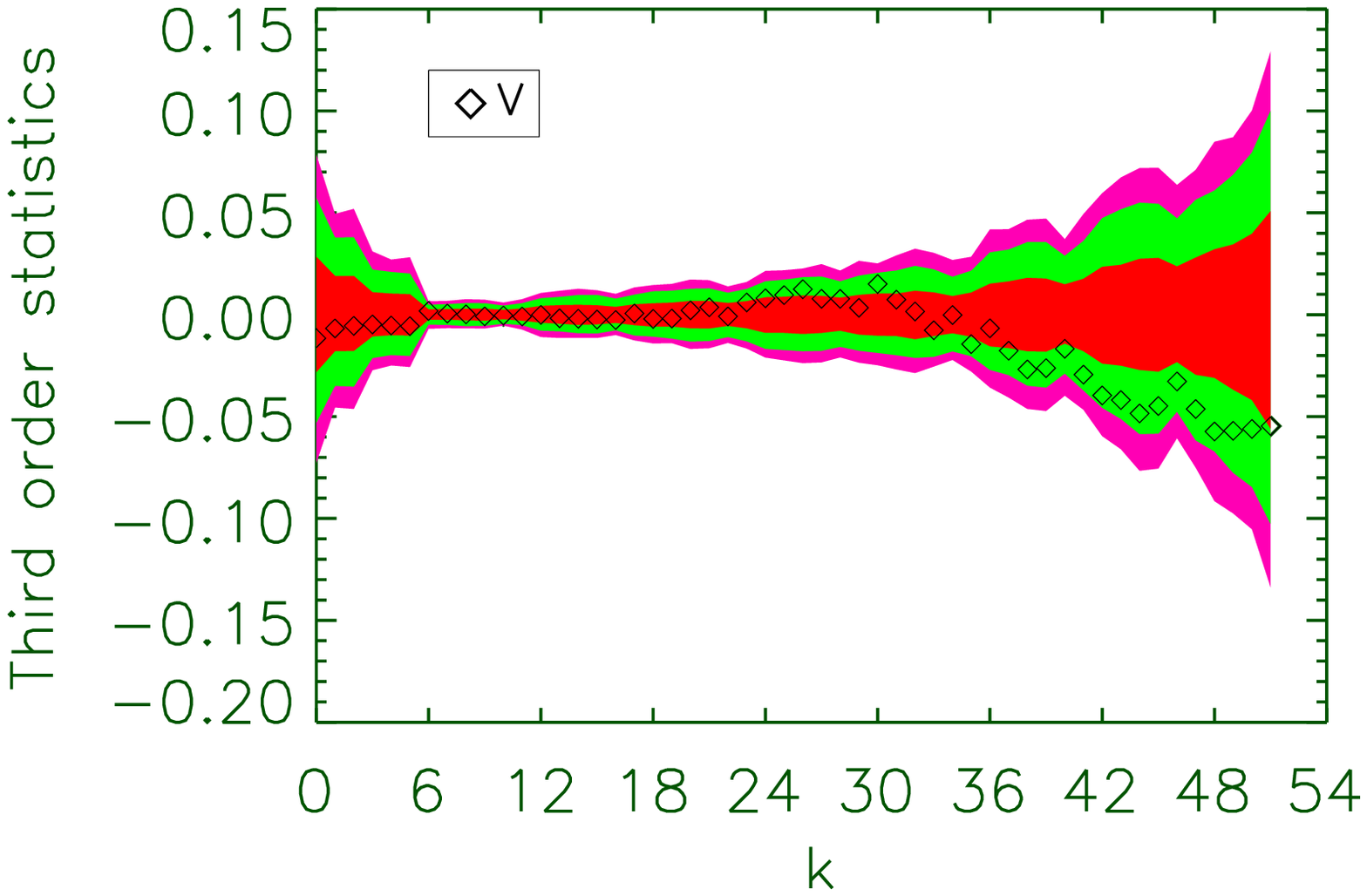}
%~~
\includegraphics[width=7.0cm,height=4.4cm] {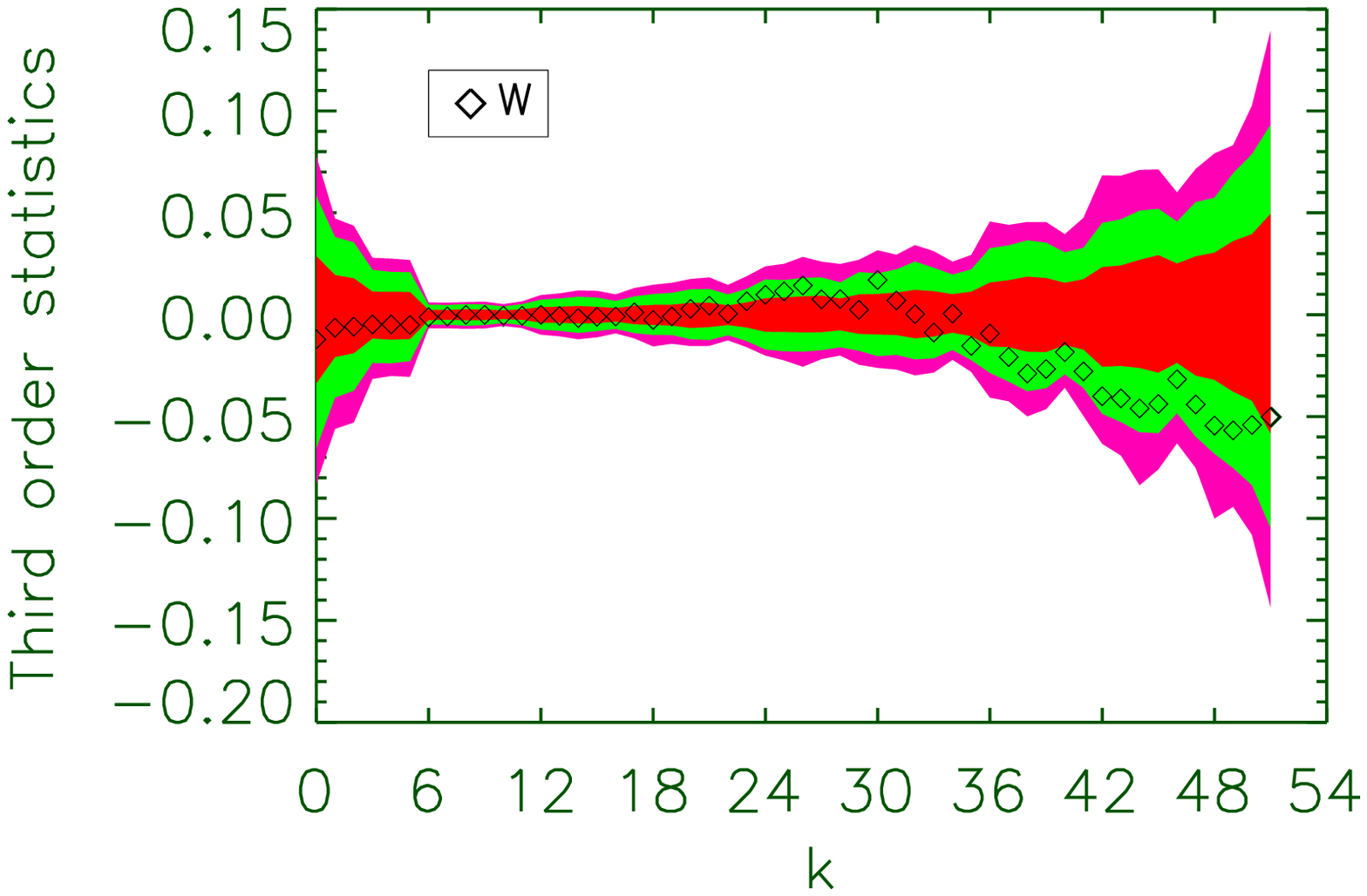}
%~~
\includegraphics[width=7.0cm,height=4.4cm] {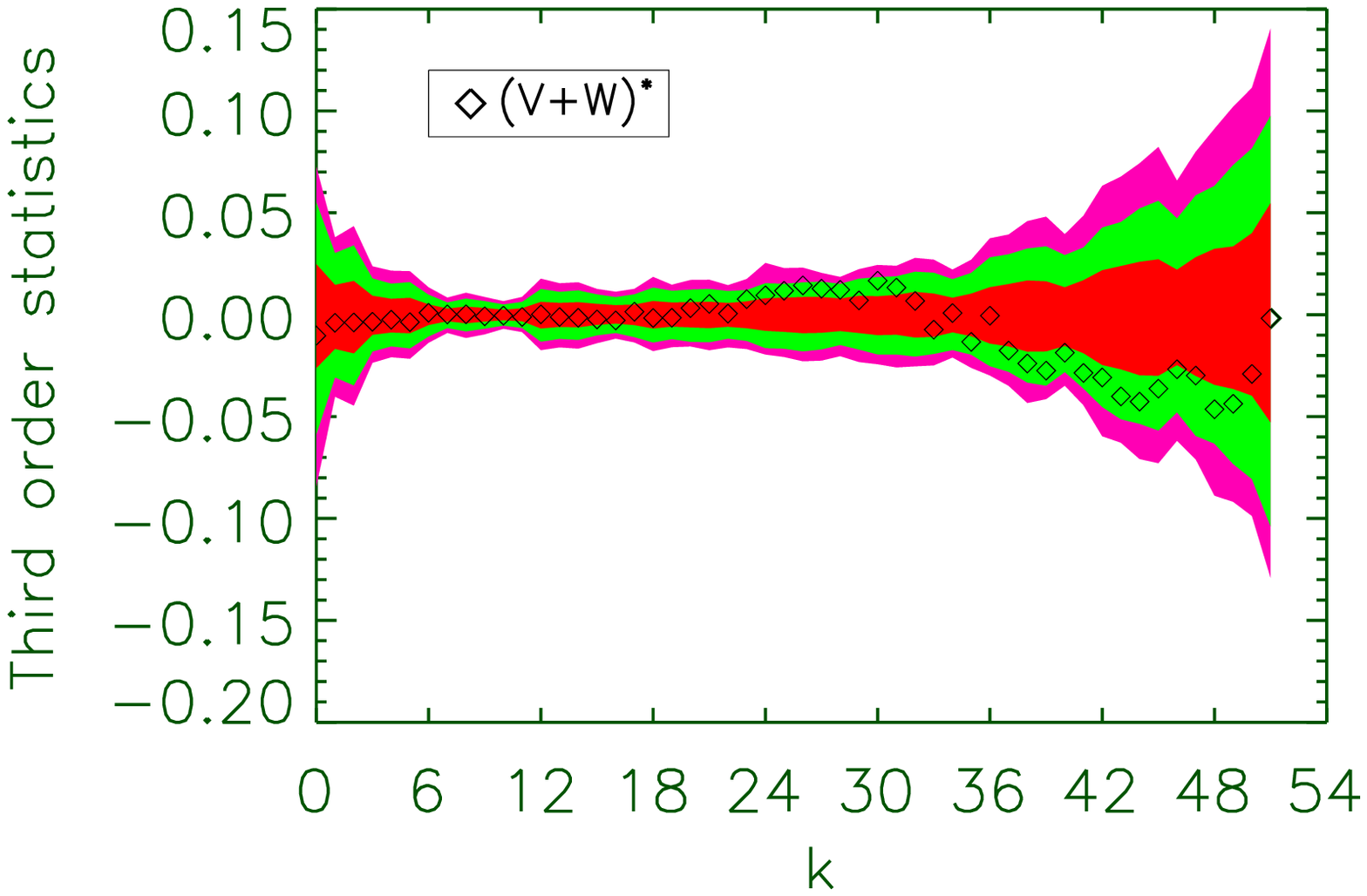}
%~~
\caption{The 52 statistics $v_k$, given in Eq. \ref{myvector},
  evaluated at 10 different scales for the WMAP combined maps. From
  left to right and top to bottom, we present the values corresponding
  to the Q+V+W, V+W, V+W, Q, V, W maps using the restrictive
    extended masks, and the V+W map using the less restrictive
    extended masks. The diamonds correspond to the data. We also plot
  the acceptance intervals for the 68\% (inner in red), 95\% (middle
  in green), and 99\% (outer in magenta) significance levels given by
  1000 Gaussian simulations of signal and noise.  }
\label{estimators_qvw_vw}
\end{figure*}
\begin{figure*}
\center
\includegraphics[width=7.0cm,height=4.4cm] {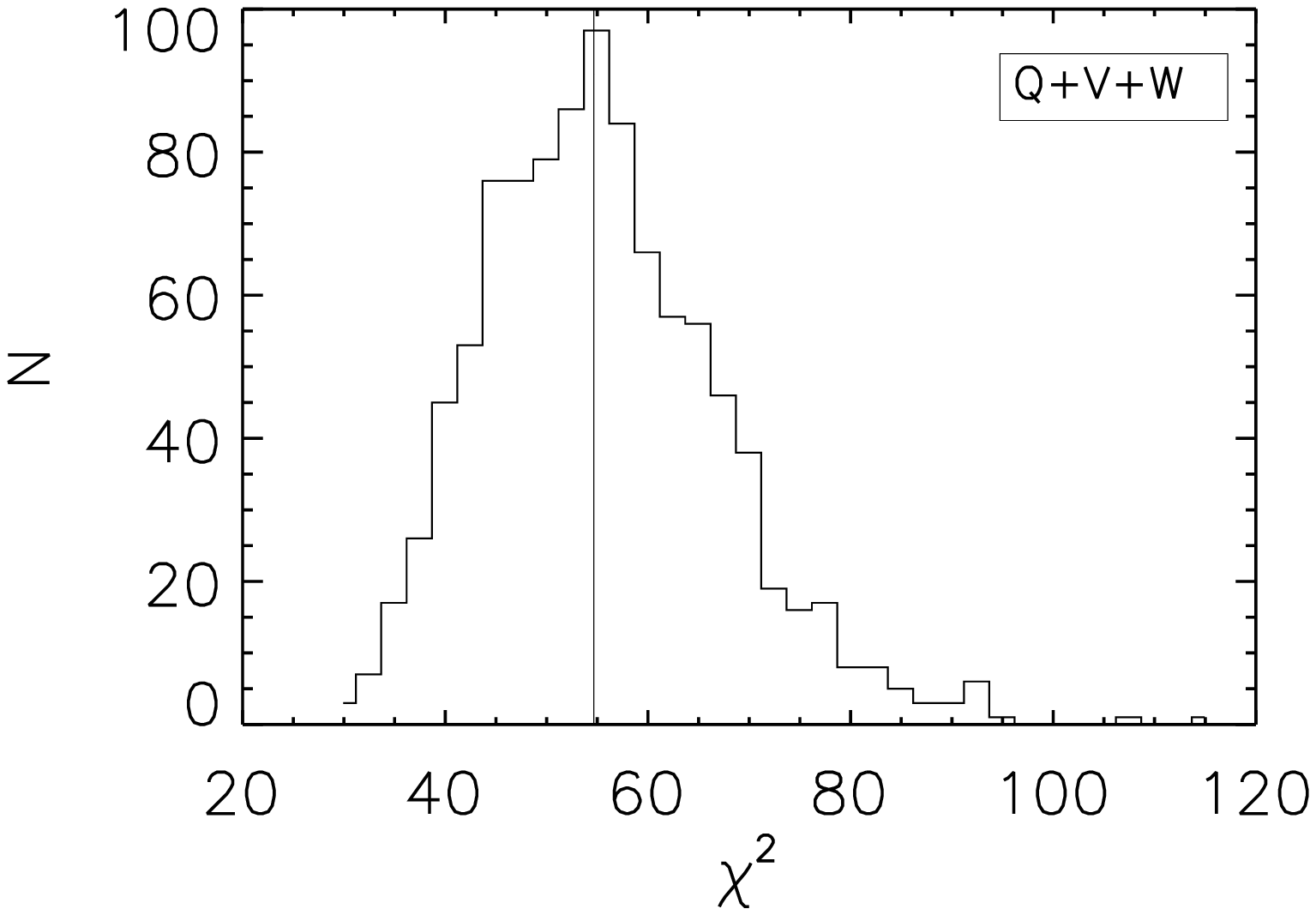}
%~~
\includegraphics[width=7.0cm,height=4.4cm] {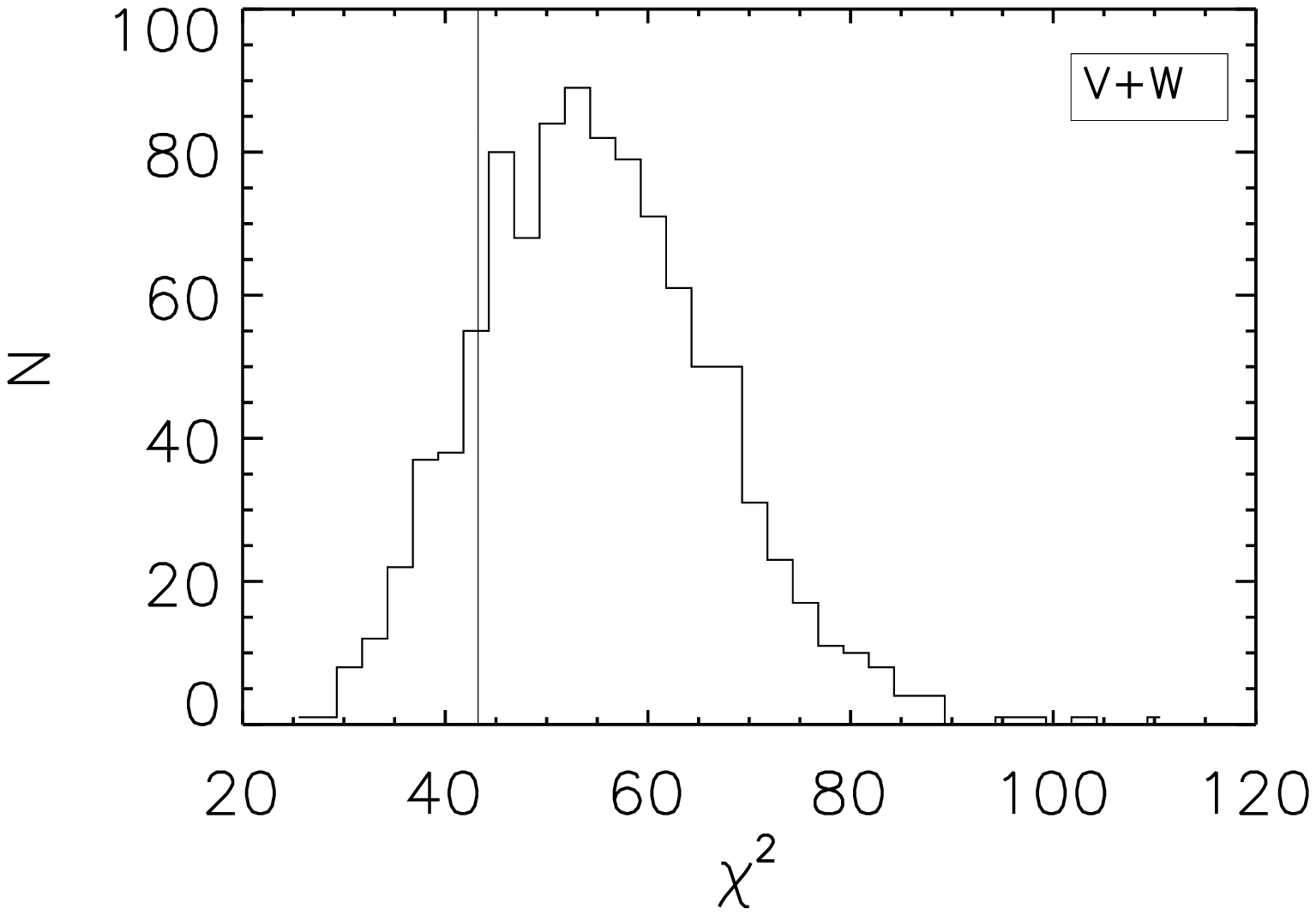}
%~~
\includegraphics[width=7.0cm,height=4.4cm] {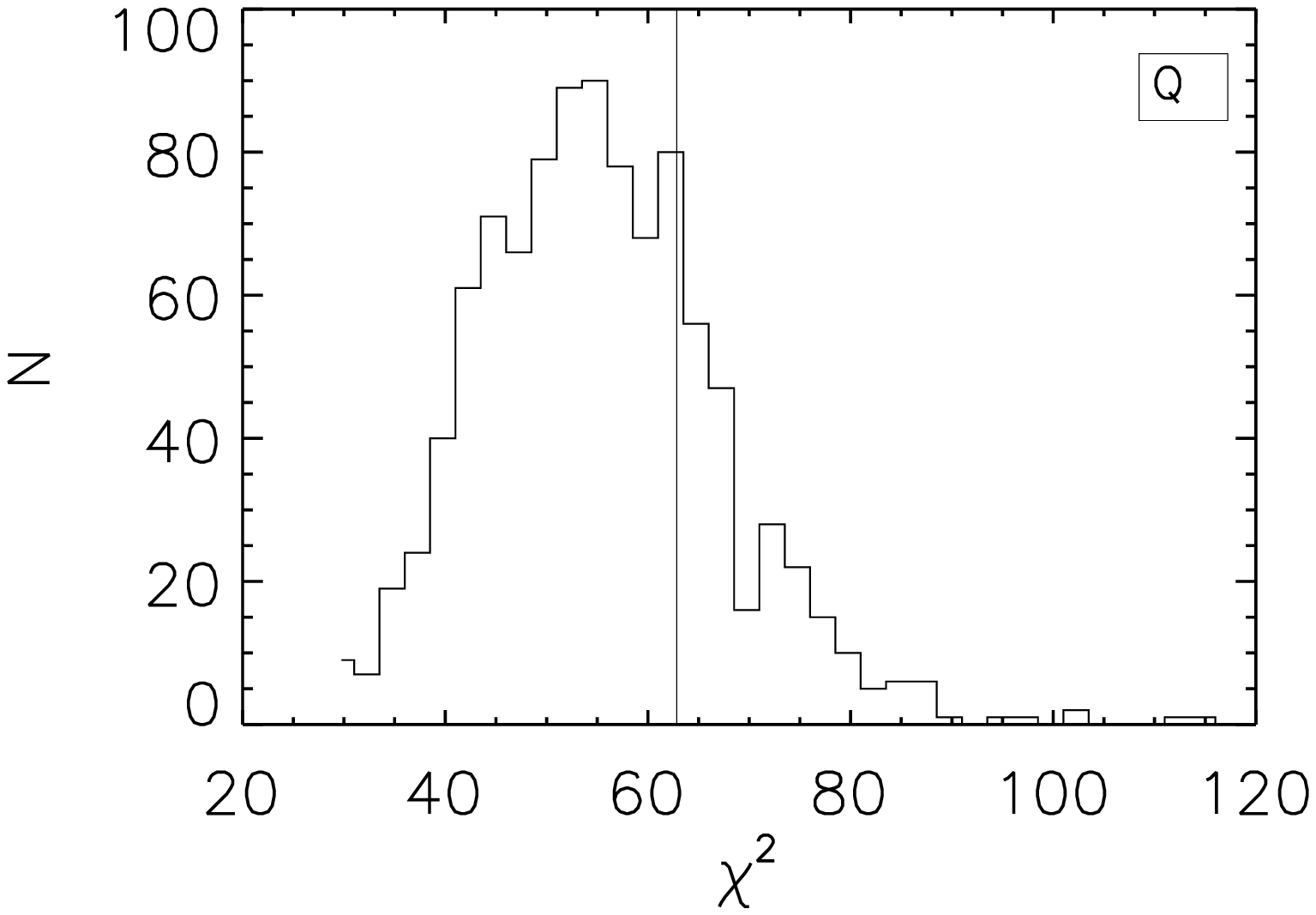}
%~~
\includegraphics[width=7.0cm,height=4.4cm] {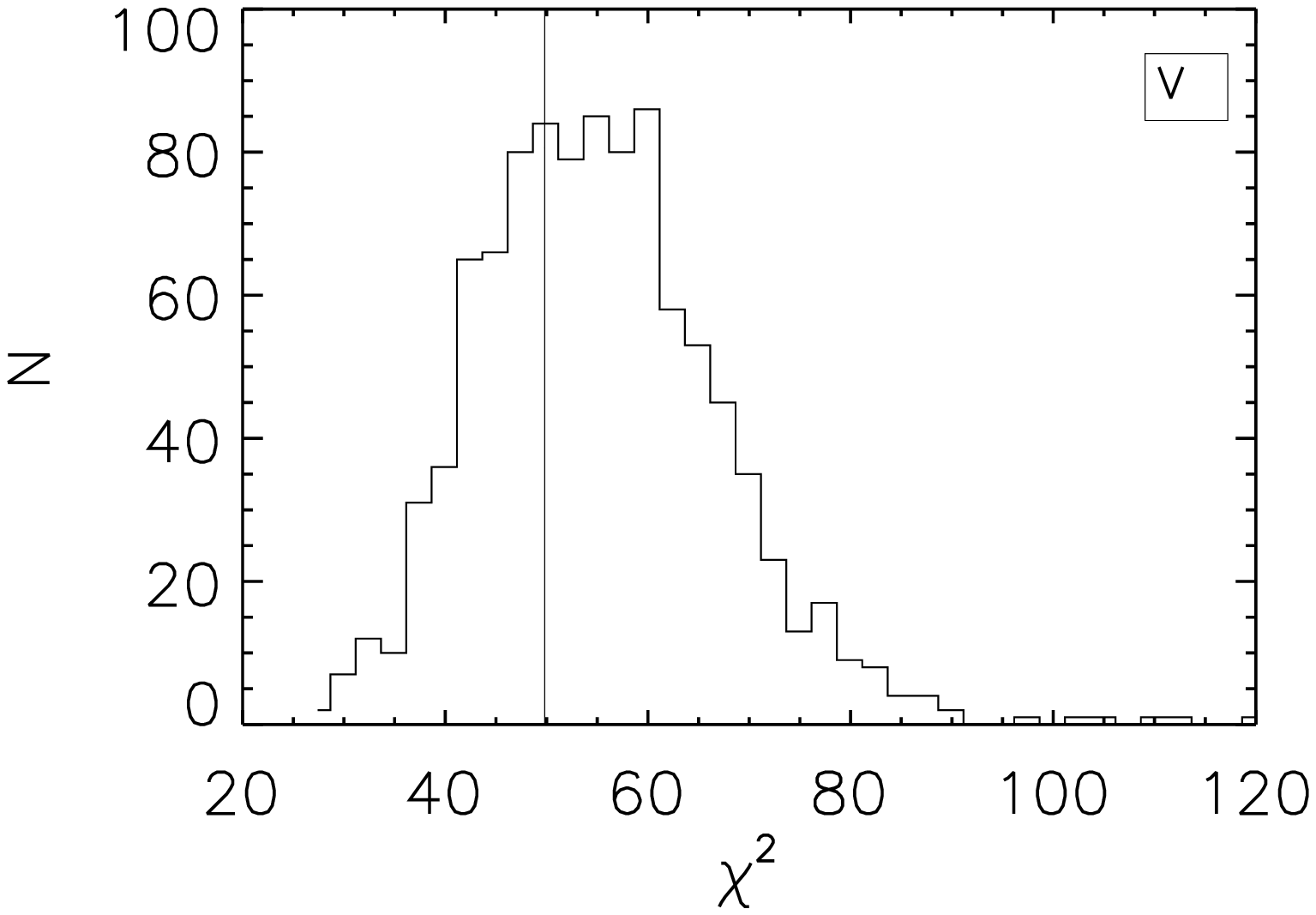}
%~~
\includegraphics[width=7.0cm,height=4.4cm] {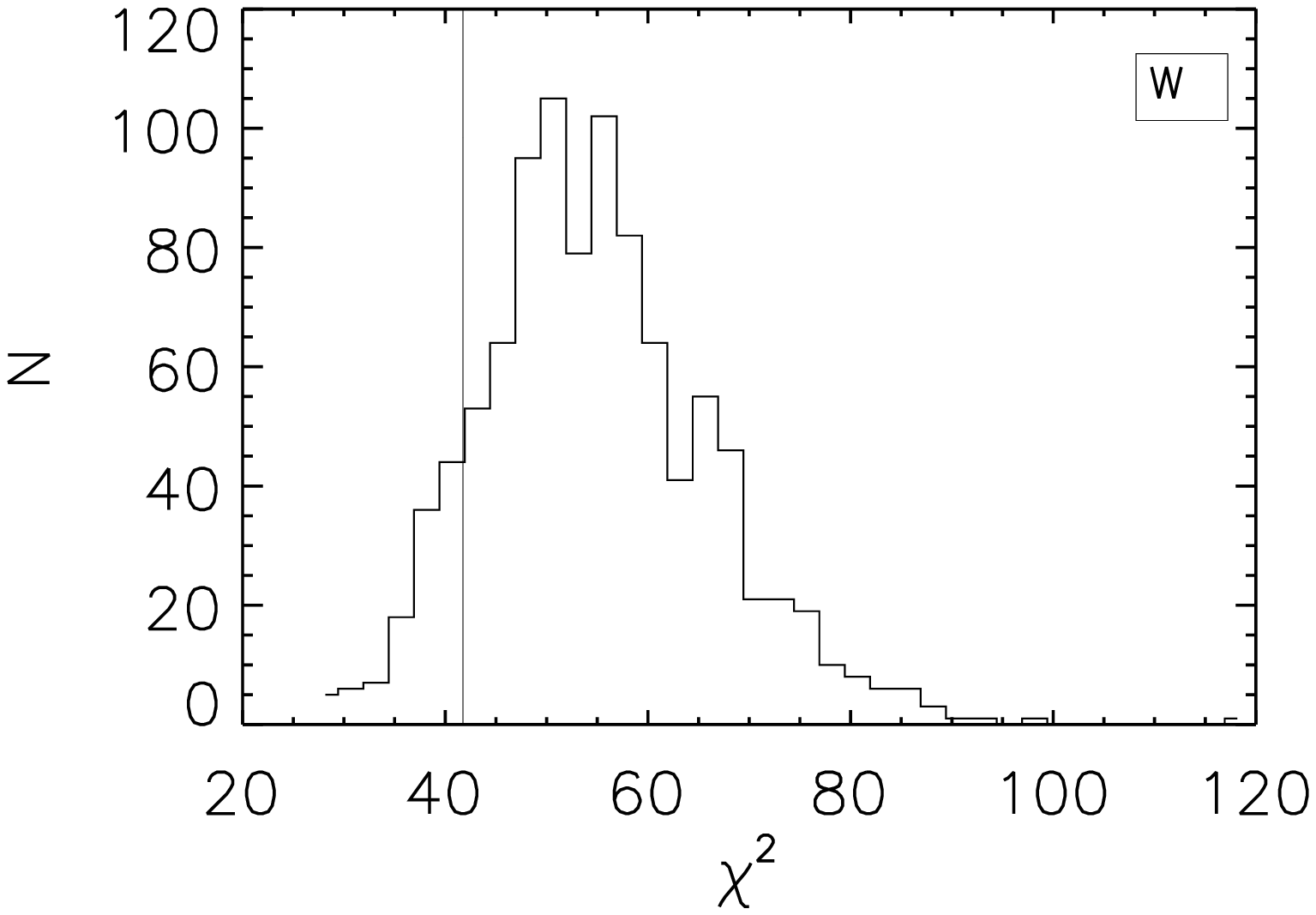}
%~~
\includegraphics[width=7.0cm,height=4.4cm] {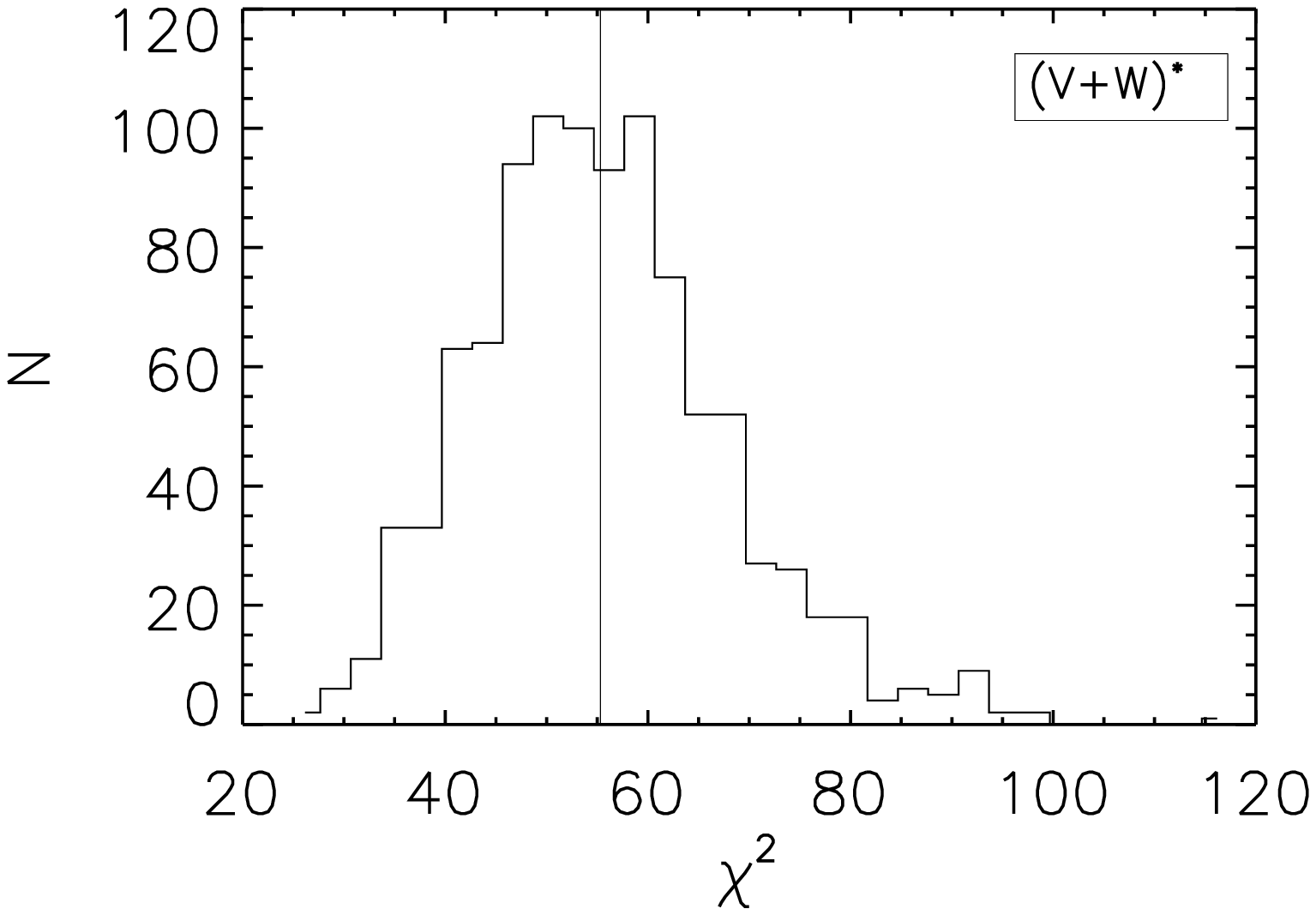}
%~~
\caption{The distribution of the $\chi^2$ statistics obtained from
  Gaussian simulations of the Q+V+W, V+W, Q, V, W maps using the
  restrictive extended masks, and the V+W map using the less
  restrictive extended masks. The vertical lines correspond to the
  values obtained from the data. }
\label{chi_qvw_vw}
\end{figure*}
We estimate the wavelet coefficient maps at the 10 considered scales
defined in Subsect.\ref{theestimators}. In particular, for $n_{sc}=
10$ we have 52 estimators as the ones defined in
Eq. \ref{witowip1towip2}. In Fig. \ref{estimators_qvw_vw} we present
these statistics for the WMAP data compared with the values obtained
for 1000 Gaussian simulations. We analyse the Q+V+W, V+W, Q, V, and W
combined maps using the restrictive extended masks, and we also
analyse the V+W map with the less restrictive extended masks (marked
in the tables and figures with $(V+W)^*$). The $q_1$ estimator
(defined in Eq. \ref{witowip22}, corresponding to the skewness) has a
similar shape than the one obtained for the 1-year and 3-year data
maps \citep{vielva2004,mukherjee2004,cruzb}. In all the cases the
skewness obtained for the data is compatible with the skewness of
Gaussian simulations. We can see in Fig. \ref{estimators_qvw_vw} that
the values of the estimators with $ k \ge 36$ are systematically below
the mean. However, for a correct interpretation of this trend it is
important to take into account that those estimators are strongly
correlated at large scales. In particular, considering the normalised
correlation matrix in the V+W case, $ C_{ij}/(\sigma_i\sigma_j)$, we
have that its elements are $ C_{ij}/(\sigma_i\sigma_j)\sim0.7$ on
average for vector indices greater or equal than 36, whereas $
C_{ij}/(\sigma_i\sigma_j)\sim0.4$ for vector indices lower than 36.

We have also performed a $\chi^2$ analysis in order to check if the
data are compatible with Gaussian simulations. This is plotted in
Fig. \ref{chi_qvw_vw}. We estimate the mean and the covariance
  used in Eq. \ref{chigauss} with another set of 1000 Gaussian
  simulations. We can see that for all the cases, the data are
compatible with the Gaussian simulations. The statistical properties
of the histograms are presented in Table \ref{stat_vw_qvw}, where we
give the $\chi^2$ value obtained for the data, the degrees of freedom
of the $\chi^2$ statistic (DOF), the mean and the dispersion obtained
for Gaussian simulations and the cumulative probability for the data.

\begin{figure*}
\center
\includegraphics[width=5.5cm,height=3.8cm] {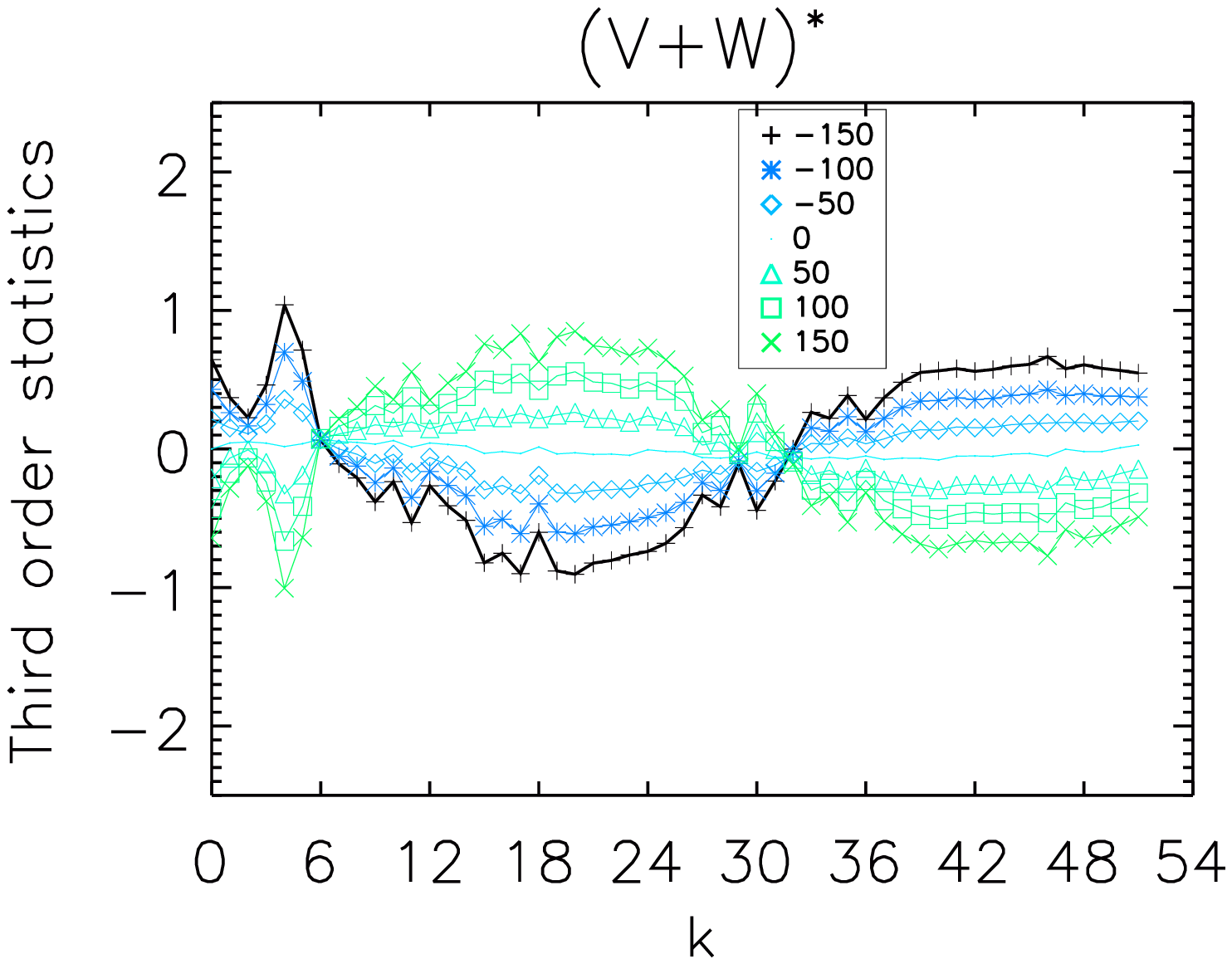}
%~~
\includegraphics[width=5.5cm,height=3.8cm] {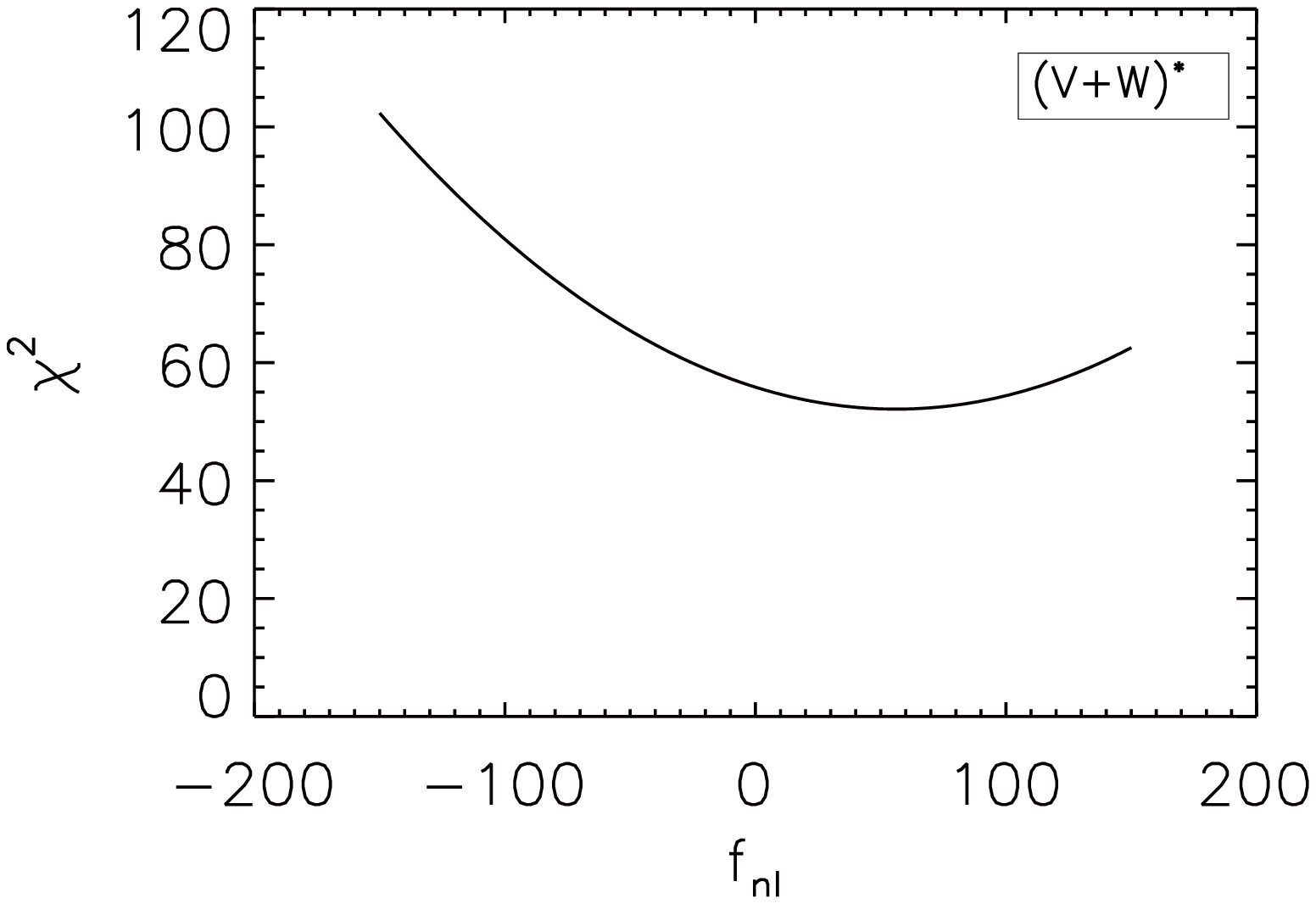}
%~~
\includegraphics[width=5.5cm,height=3.8cm] {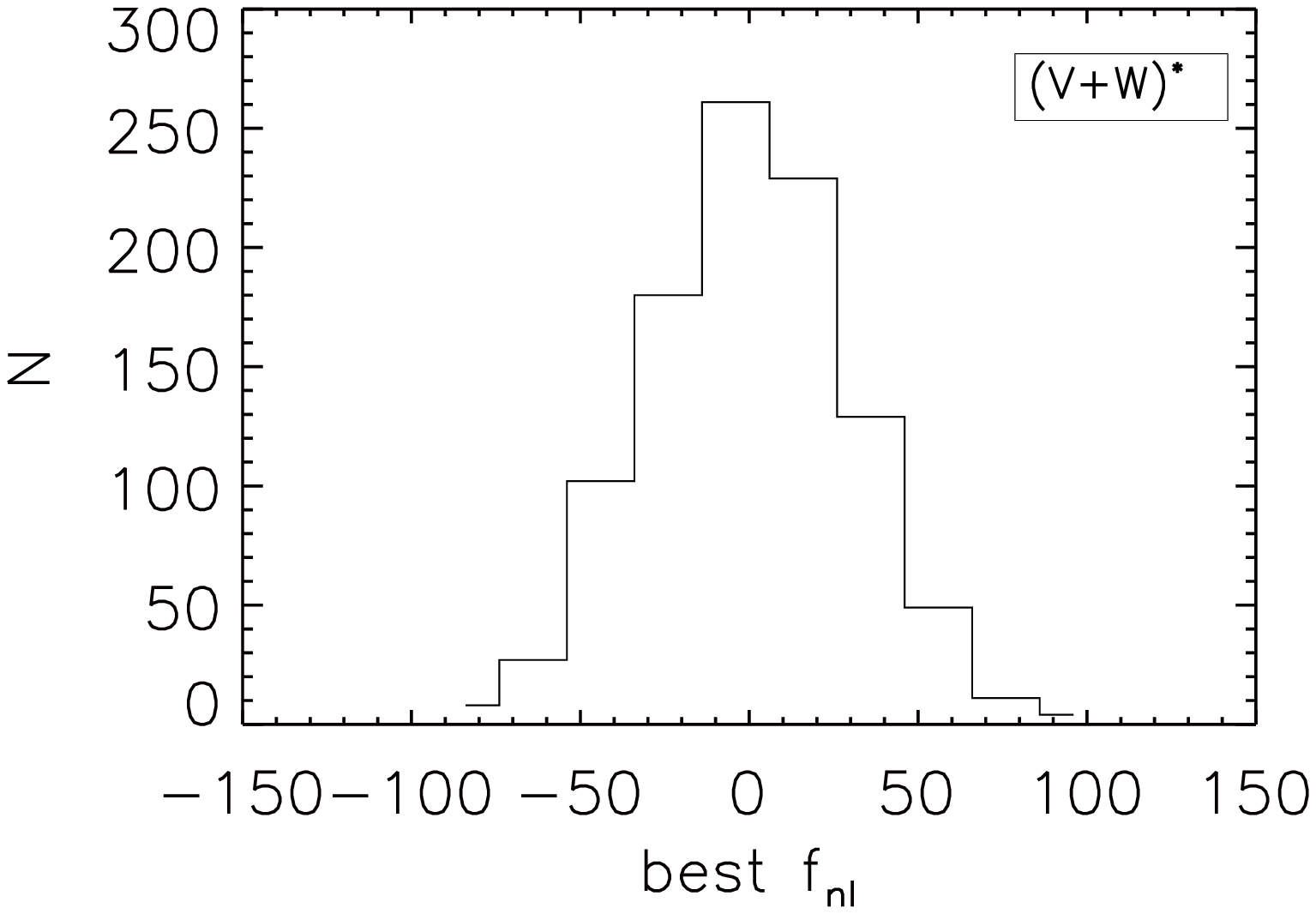}
%~~
\caption{From left to right, the normalised mean values of the
  estimators for 300 simulations of the V+W map with different
  $f_{nl}$ contributions, the $\chi^2(f_{nl})$ statistics for the WMAP
  combined V+W data map, and the histogram of the best fit $f_{nl}$
  values for a set of 1000 Gaussian simulations of the V+W map.  We
  have used the less restrictive extended masks. Similar results are
  obtained with the restrictive extended masks.}
\label{best_fnl_vw}
\end{figure*}

Finally we impose constraints on the $f_{nl}$ parameter using the
considered maps. We calculate the expected values of all the
estimators for different $f_{nl}$ cases\footnote{ We use 300
    simulations to calculate the mean values of the estimators for
    different $ f_{nl}$ cases and 1000 Gaussian simulations to
    calculate the covariance matrix. We analysed the convergence of
    the mean values by computing the best-fit $ f_{nl}$ value of
    the data using two independent sets of 150 simulations. The
    difference in the obtained value is lower than 1 and therefore
    even with $ \sim$100 simulations the convergence is
    achieved. Analogously, for the covariance matrix we have checked
    that convergence is achieved with $ \sim$1000 Gaussian
    simulations.} (see left panel of Fig. \ref{best_fnl_vw} for the
combined V+W map). Then, we perform a $\chi^2$ analysis using
Eq. \ref{chifnl} in order to find the best fit value for $f_{nl}$ for
each data map. We also analyse Gaussian simulations in order to obtain
the frequentist error bars. In the central and right panels of
Fig. \ref{best_fnl_vw} we present a plot of $\chi^2(f_{nl})$ versus
$f_{nl}$ for the combined V+W data and a histogram of the best fit
$f_{nl}$ for 1000 Gaussian simulations of the combined V+W map. Table
\ref{bestfnl_vw_qvw} lists the $f_{nl}$ values which best fits the
five considered maps, and the main properties of the distributions of
best-fit $f_{nl}$ obtained from Gaussian simulations. From this table
we have at 95\% CL that $-8 < f_{nl} < 118$ for the combined V+W map,
$-42 < f_{nl}< +93$ for Q+V+W, $-63 < f_{nl}< +102$ for Q, $-46 <
f_{nl}< +109$ for V, and $-38 < f_{nl}< +114$ for W (using the
restrictive extended masks defined in Subsect. \ref{datasimu}). These
values are compatible with the ones obtained by \citet{komatsu2008}
using the bispectrum. Notice that $f_{nl}$ increases as we go from Q
to W maps. The $f_{nl}$ value obtained for the V+W map is the same as
the one that \citet{komatsu2008} obtains for this map using
$\ell_{max}=700$ and the $KQ$75 mask. The Q+V+W map has a smaller
$f_{nl}$ value. The best-fit $f_{nl}$ increases with frequency for the
Q, V, and W maps, and the $\sigma(f_{nl})$ of the Q map is larger than
the one of the V and W maps. This also agrees with
\citet{komatsu2008}.

The less restrictive extended masks give tighter results for the
constraints on $f_{nl}$ compared with the restrictive ones. In
particular, from Table \ref{bestfnl_vw_qvw}, we have that $-5 <
f_{nl}< +114$ at 95\% CL. In this case the dispersion of $f_{nl}$ is
smaller since the available area is larger for all the scales. The
additional pixels are not significantly affected by the mask for this
analysis and the wavelet is still efficient. These constraints are
very similar to those obtained with the bispectrum
\citep{komatsu2008}, which are $-5 < f_{nl} < +115$ at 95\% CL.

Notice however that other works, e.g.  \citet{wandelt2008}, have found
evidences of non-zero value for $f_{nl}$ at $ 2.8\sigma$ CL using the
KSW bispectrum. We do not find deviations with respect to the zero
value using the 5-year data at 95 \% CL and the same result is
presented in \citet{komatsu2008}. We may wonder if that deviation is
introduced by the 3-year WMAP mask (Kp0), uncertainties in the
characterisation of the beams, or some systematics present in the
3-year data. To check this possibility we analysed the V+W 3-year WMAP
data (updated after the 5-year release) using our estimators, the Kp0
mask and its corresponding extended masks. The results are $ -17 <
f_{nl} < +108$ at 95\% CL, which are compatible with the 5-year
results. Therefore our method is not able to detect the deviations
reported in \citet{wandelt2008} using the 3-year data. This may be
explained by the fact that the wavelet method probes different
combinations of scales than the bispectrum and also responds
differently to possible systematics present in the data.

We have also tested other third order estimators that involve the
derivatives (in particular the squared modulus of the gradient and the
Laplacian) of the wavelet coefficient maps. These estimators are less
sensitive to the $f_{nl}$ parameter and its error bars are
significantly wider than the ones obtained with the estimators of
Eq. \ref{witowip22}.
\begin{table*}
  \center
  \caption{$\chi^2$ constructed from the 5statistics for Q+V+W,
    V+W, Q, V, and W data maps using
    Eq. \ref{chigauss}. \label{stat_vw_qvw} We also present the mean
    and the dispersion of the $\chi^2$ corresponding to 1000 Gaussian
    simulations, and the cumulative probability for the $\chi^2$ of
    the data.}
  \begin{tabular}{cccccc}
    \hline 
    \hline
    Map & $\chi_{data}^2$ & DOF & $\langle \chi^2 \rangle$ & $\sigma$ & $P(\chi^2\le \chi_{data}^2)$\\
    \hline
    Q+V+W & 54.65 & 52 & 55.49 & 11.78 & 0.63 \\
      V+W & 43.23 & 52 & 55.22 & 11.73 & 0.20 \\
      Q & 62.84 & 52 & 55.31 & 12.08 & 0.86 \\
      V & 49.81 & 52 & 55.24 & 11.96 & 0.44 \\
      W & 41.74 & 52 & 54.92 & 11.34 & 0.16\\
      (V+W)$^*$ & 55.31 & 52 & 55.47 & 12.47 & 0.65 \\
    \hline
    \hline
  \end{tabular}
\flushleft
 $^*$ The V+W map is analysed using the less restrictive extended masks.
\end{table*}
\begin{table*}
  \center
  \caption{Best fit $f_{nl}$ values obtained from Q+V+W, V+W, Q, V
    and W combined maps. We also present the mean, dispersion and some
    percentiles of the distribution of the best fit $f_{nl}$ values
    obtained from Gaussian simulations. \label{bestfnl_vw_qvw}}
  \begin{tabular}{cc|cc|cc|cc}
    \hline 
    \hline
    Map & best $f_{nl}$ & $\langle f_{nl} \rangle$ & $\sigma(f_{nl})$ &  $X_{0.160}$ & $X_{0.840}$ & $X_{0.025}$ & $X_{0.975}$ \\
    \hline
    Q+V+W & 26. & -2. & 34. & -34. & 30. & -68. & 67. \\
      V+W & 58. & -1. & 33. & -35. & 34. & -66. & 60. \\
      Q & 16. & -1. & 42. & -43. & 41. & -79. & 86. \\
      V & 30. & 0. & 39. & -39. & 38. & -76. & 79. \\
      W & 40. & -4. & 39. & -44. & 35. & -78. & 74. \\
      (V+W)$^*$ & 56. & 0. & 30. & -31. & 30. & -61. & 58. \\
    \hline
    \hline
  \end{tabular}
\flushleft
$^*$ The V+W map is analysed using the less restrictive extended masks.
\end{table*}
\subsection{Contribution of undetected point sources}
It has been shown that the Mexican hat wavelet is very useful in
detecting point sources \citep{cayon2000,vielva2001} since its
contribution is enhanced in the wavelet coefficient maps at certain
scales. Therefore it is important to study the contribution of the
undetected point sources on the $f_{nl}$ estimation. The skewness and
kurtosis of wavelet coefficient maps due to point sources have been
studied in \citet{argueso2006}. We estimate the contribution of the
point sources to our estimators by performing Monte Carlo simulations.
We use a straightforward model to simulate the radio sources following
the one described in \citet{komatsu2008}. In this model it is assumed
that all the sources have the same intensity flux, ($F_{src}=$0.5 Jy),
and that they are randomly distributed in the sky following a
Poisson distribution. The source contribution to the temperature at
any pixel is given by
\begin{equation}
T_{src}({\bf n}) = \frac{\sinh(x/2)^2}{x^4}\frac{1}{24.8~MJy/K}\frac{F_{src}}{\Omega_{pix}}\epsilon
\label{pssimus}
\end{equation}
where $x=\nu/(56.8~GHz)$, $\Omega_{pix}=4\pi/N_{pix}$, 
$\epsilon$ is a Poisson random variable of mean $\langle \epsilon
\rangle = \Omega_{pix}n_{src}$ and $n_{src}$ is the density of
sources. The power spectrum for this type of source distribution can
be easily calculated \citep{tegmark96}.
\begin{equation}
C_l^{ps}=n_{src}T_0^2\frac{\sinh^4(x/2)}{x^8}\left(\frac{F_{src}}{67.55~MJy}\right)^2
\end{equation}
Using $n_{src}~=~85~sr^{-1}$ and a pixel resolution of $N_{side}=512$
we have a power spectrum of $C_l^{ps}~=~8.68 \times 10^{-3}$ $\mu
K^2sr$ for the Q band. This model roughly reproduces the measured
values of the power spectrum and the bispectrum
\citep{nolta2008,komatsu2008}.

\begin{figure*}
\center
\includegraphics[width=7.0cm,height=4.4cm] {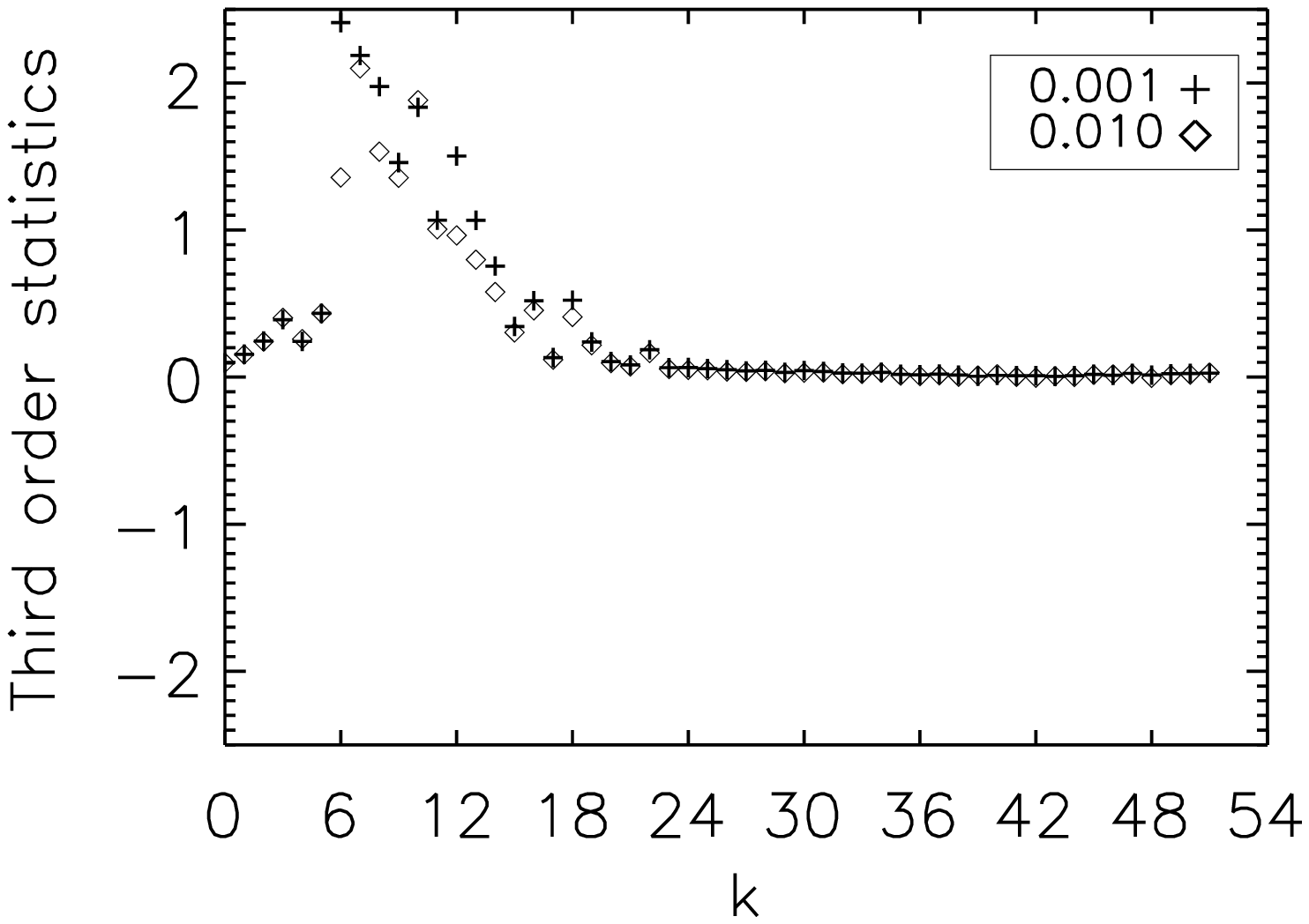}
\includegraphics[width=7.0cm,height=4.4cm] {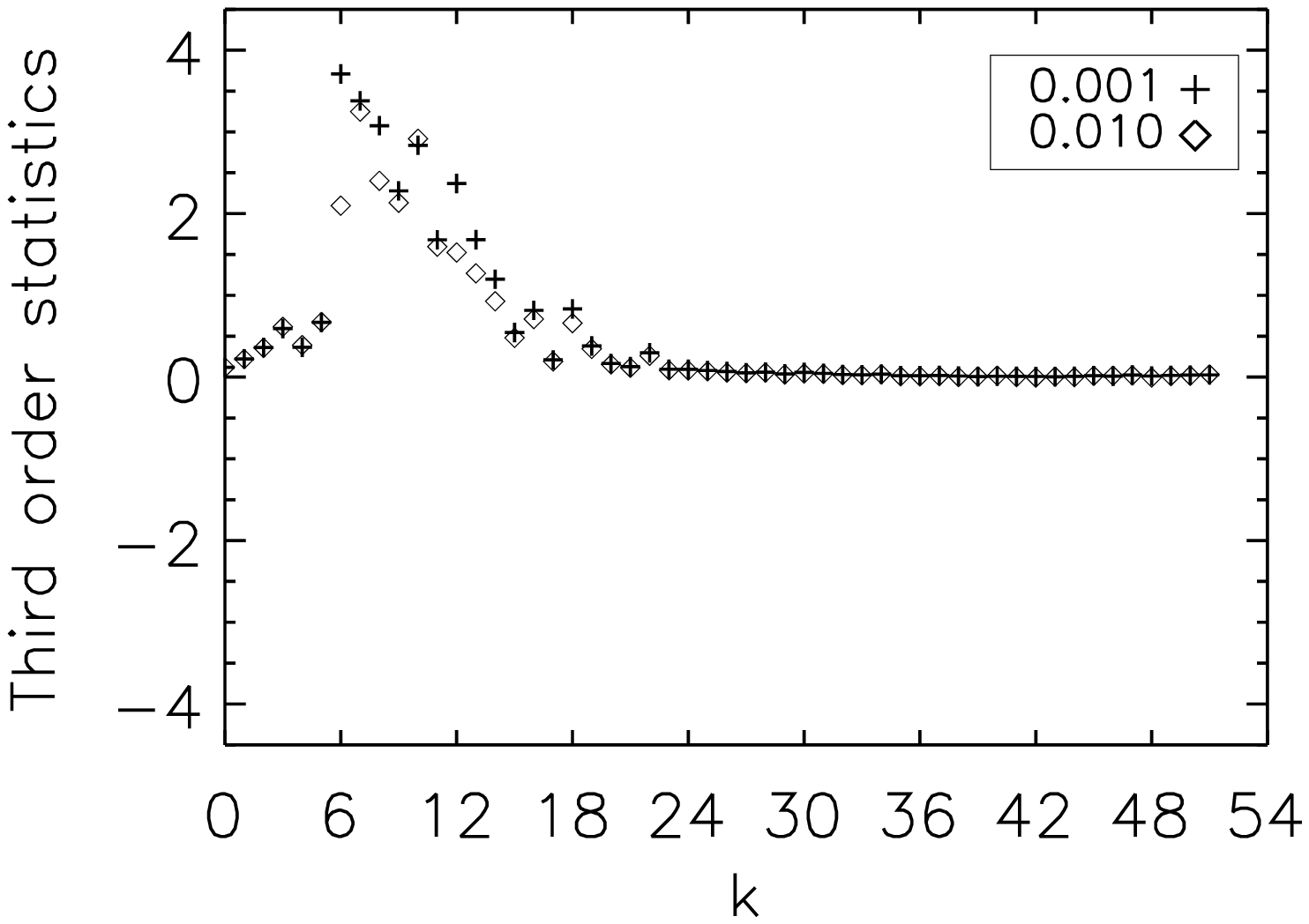}
%~~
%~~
\caption{Normalised mean value of the third order estimators for 1000
  Gaussian simulations plus point source simulations of the V+W
  combined map using the restrictive mask (threshold of 0.001) and the
  less restrictive mask (threshold of 0.010). In the left panel
    we use the model with point sources of constant flux, $
    F_{src}$=0.5 Jy, and in the right panel we use the model with a
    $ dN/dS$ derived from \citet{zotti}.}
\label{pointsources_vw}
\end{figure*}

For this analysis we have performed 1000 point source simulations for
the V+W map. We add these point source simulations to the V+W
simulations of CMB with noise.  For each one we estimate its best fit
$f_{nl}$ and compare it with the values obtained for the same
simulation but without point sources. The difference gives us an
estimate of the contamination due to the point sources.  They add a
contribution of $\Delta f_{nl}=11 \pm 4$ to the CMB considering the
restrictive extended masks. For the less restrictive extended masks
the point source contribution is $\Delta f_{nl} = 3 \pm 4$. The
difference with the previous result is explained because the zero
values of the $KQ$75 mask affect the efficiency of the wavelet to
detect point sources. This can be seen in the left panel of 
Fig.  \ref{pointsources_vw} where we plot the mean values for the
third order estimators for the V+W simulations including point sources
for both kinds of extended masks. The mean value of several
estimators, specially the skewness for small scales, is lower for the
less restrictive extended masks. This reduces the $\Delta f_{nl}$
value. Considering the contribution of the point sources, our estimate
of $f_{nl}$ for this case is $+22 < f_{nl} < +83$ at 68\% CL and $-8 <
f_{nl} < +111$ at 95\% CL. For comparison, the bias introduced by
point sources for the bispectrum is $\Delta f_{nl} = 5 \pm 2$
\citep{komatsu2008}.

We also estimate the contribution of undetected point sources using
the more realistic source number counts $ dN/dS$, derived from the
work of \citet{zotti}. The dependence of $ dN/dS$ with the frequency
is very small in the range between 61 GHz and 94 GHz. Thus, for
simplicity, we assume the same $ dN/dS$ for both V and W maps,
evaluated at a frequency of 71 GHz. We select a range of intensities
between $ S_{min}=$ 1 mJy and $ S_{max}=$ 1 Jy. Then we generate point
source simulations following the distribution $ dN/dS$ derived from
\citet{zotti}. The simulations are transformed from intensity flux to
temperature as in the previous case (following Eq. \ref{pssimus}). For
this analysis we have also performed 1000 point source simulations for
the $ V+W$ map and have added them to Gaussian CMB plus noise
simulations. We analyse the resultant simulations as in the previous
case to obtain the contribution to $ f_{nl}$ due to the point
sources. They add a contribution of $ \Delta f_{nl}= 3 \pm 5$ to the
CMB considering the less restrictive extended masks and $ \Delta
f_{nl}= 17 \pm 5$ considering the restrictive extended masks. As in
the previous case, the differences in $ \Delta f_{nl}$ using the
restrictive and less restrictive extended masks are explained because
the zero values of the $ KQ$75 mask affect the efficiency of the
wavelet. This is plotted in the right panel of Fig.
\ref{pointsources_vw}.

Therefore, considering the realistic point source model and the less
restrictive masks our estimate of $f_{nl}$ remains unchanged ($ -8 <
f_{nl} < +111$ at 95\% CL).
\section{Conclusions}
\label{conclusions}
We have analysed the 5-year WMAP data using Gaussian and realistic
non-Gaussian simulations through a wavelet-based test. We have
considered different combined maps for the analysis: Q+V+W, V+W, Q, V,
and W. We have considered two kinds of extended masks to analyse the
V+W map (defined in Subsect. \ref{datasimu}). The third order moments
(Eq. \ref{witowip22}) of the wavelet coefficients of these maps are
compatible with Gaussian simulations (see Fig
\ref{estimators_qvw_vw}).

We have performed a $\chi^2$ analysis and found that the data are
indeed compatible with Gaussian simulations (see
Fig. \ref{chi_qvw_vw}). We performed another $\chi^2$ analysis to
constrain the non-linear coupling parameter $f_{nl}$ by using
non-Gaussian simulations with $f_{nl}$. The best-fit $f_{nl}$ values
of the five analysed maps are compatible with the ones obtained by
\citet{komatsu2008} using the bispectrum, showing similar confidence
intervals and also a similar trend with the frequency.

Finally we have estimated the contribution to $f_{nl}$ from unresolved
point sources for the V+W map using a simple model that has point
  sources with constant intensity ($F_{src}=$ 0.5 Jy) and a realistic
  model given by \citet{zotti}. We have found that they add a
positive contribution of $\Delta f_{nl} = 11 \pm 4$ for the
  simple model and $ \Delta f_{nl} = 17 \pm 5$ for the realistic
  model. These values are larger than the one obtained by
\citet{komatsu2008} for the bispectrum, which can be explained by the
enhancement of the point sources produced by wavelets. Using the less
restrictive extended masks, the point source distribution add a
positive contribution of $\Delta f_{nl} = 3 \pm 4$ for the simple
  model and $ \Delta f_{nl} = 3 \pm 5$ for the realistic
  model. The smaller values are explained because the less
restrictive extended masks add some pixels that affect the efficiency
of the wavelet to detect point sources.  Taking into account the point
source correction, our best estimate of $f_{nl}$ is $-8 < f_{nl} <
+111$ at 95\% CL.  It is important to emphasise the agreement found
between the two estimators (bispectrum and wavelets), since they are
formed by very different combinations of the data and are affected by
systematic effects (like the mask, noise, beam response and foreground
residuals) in very different ways.
\section*{acknowledgments}
The authors thanks J. Gonz\'alez-Nuevo for providing the $ dN/dS$
counts and for his useful comments on unresolved point sources. The
authors thank P. Vielva and M. Cruz. We also thank R. Marco and
L. Cabellos for computational support. We acknowledge partial
financial support from the Spanish Ministerio de Ciencia e
Innovaci\'on project AYA2007-68058-C03-02. A. C. thanks the Spanish
Ministerio de Ciencia e Innovaci\'on for a pre-doctoral fellowship.
A. C. thanks the Astronomy Centre at the University of Sussex for
their hospitality during a research stay in 2008. S. M. thanks ASI
contract I/016/07/0 "COFIS" and ASI contract Planck LFI activity of
Phase E2 for partial financial support. The authors acknowledge the
computer resources, technical expertise and assistance provided by the
Spanish Supercomputing Network (RES) node at Universidad de
Cantabria. We acknowledge the use of Legacy Archive for Microwave
Background Data Analysis (LAMBDA). Support for it is provided by the
NASA Office of Space Science. The HEALPix package was used throughout
the data analysis \citep{healpix}.
%
%
%\bibliographystyle{mn2e}
%\bibliography{mybib}
%

%
\end{document}